\documentclass{aa}

\usepackage[utf8]{inputenc}

\usepackage{hyperref}
\usepackage{graphicx}
\usepackage{amsmath}
\usepackage{amssymb}
\usepackage{txfonts}
\usepackage[normalem]{ulem}

\begin{document}

\title{Warm and thick corona for magnetically supported disk in GBHB}

\author{D.~Gronkiewicz\inst{1}
       \and A.~R\'o\.za\'nska \inst{1}}
   \offprints{D.~Gronkiewicz}

\institute{Nicolaus Copernicus Astronomical Center, Polish Academy of Sciences, Bartycka 18,
           00-716 Warsaw, Poland \\
            \email{gronki@camk.edu.pl}
           }

\date{Received September 15, 1996; accepted March 16, 1997}

\abstract
{This paper is devoted to self-consistent modeling of the magnetically supported 
accretion disk with optically thick warm corona based on first principles. In our model, we consider 
the gas heating by magneto-rotational instability (MRI) dynamo.}
{Our goal is to show that the proper calculation of the gas heating by magnetic dynamo can build up the  warm, optically thick corona above the accretion disk around black hole of stellar mass.}
{Using vertical model of the disk supported and heated by the magnetic field 
together with radiative transfer in hydrostatic and radiative equilibrium we developed 
relaxation numerical scheme which allows us to compute the transition form the disk to corona in a self consistent way.}
{We demonstrate here that the warm (up to 5 keV), optically thick (up to 10 $\tau_{\rm es}$), Compton cooled corona can form due to the magnetic 
heating. Such  warm corona is stronger for higher accretion rate and larger magnetic field strength. 
The radial extent of the warm corona is limited by the occurrence of the local thermal 
instability, which purely depends on radiative processes. The obtained coronal parameters
are in agreement with those constrained from X-ray observations.}
{The warm magnetically supported corona is tends to appear in the inner disk regions. It may be 
responsible for Soft X-ray excess seen in accreting sources. For lower accretion rates 
and weaker magnetic field parameters, thermal instability prevents warm corona to exist, 
giving rise to eventual clumpiness or ionized outflow.}


\maketitle


\section{Introduction}

There is a growing number of evidence that warm, optically thick corona exists in accreting black holes,  whenever an accretion disk is present.
It is observed in different types of accreting black holes across masses: 
active galactic nuclei (AGN) \citep{1987-Pounds-Mkn335,1998-Magdziarz,2011-Mehdipour-Mrk509,2012-Done-SXE,2013-Petrucci-Mrk509,2016-Keek,2017-Petrucci} including 
quasars \citep{1988-Madau,1994-Laor-1,1997-Laor-2,2004-GierlinskiDone,2005-Piconcelli-SXE}, ultraluminous X-ray sources (ULXs) \citep{2006-Goad,2006-Stobbart-ULX,2009-Gladstone-ULX}, and 
Galactic black holes binaries (GBHBs) \citep{1999-Gierlinski,2000-Zhang,2001-DiSalvo}.
Such warm corona is visible as a soft component (below 2  keV) presented in X-ray spectra of those objects. It is generally characterized by an excess with respect to the extrapolation of 
hard X-ray power law, the last typically originating from the region named hot corona.

The observed spectral shape of soft X-ray excess provides us crude information about radiative cooling mechanism which operates in warm corona. 
Assuming that the thermal Comptonization is the dominant cooling process, the observed spectra can be modeled  with so called {\it slab} model, where soft photons from the disk enter the warm corona located above the disk,  and undergo Compton scattering with amplification factor $y$.  
During the fitting procedure, the temperature and optical depth of a warm corona can 
be determined. The most common result from many sources is the fact that such corona 
is optically thick from 4 to 40 in different objects \citep[][and references therein]{2000-Zhang, jin12,2013-Petrucci-Mrk509,2017-Petrucci}. 

Furthermore, observations constrain the amount of energy dissipated in corona, $f$, in comparison to the total energy released in the whole disk/corona system.
Observations of both GBHBs and AGNs show that $f \approx 1$ is required to explained the observed spectral index $\alpha = 0.9$ in some sources \citep{1991-HaardtMaraschi,2001-Zycki,2013-Petrucci-Mrk509}.
\cite{2017-Petrucci} sets the lower limit of $f = 0.8$.
This behavior is not universal, as it depends on the object and its spectral state, but shows that in some circumstances a big fraction of energy is released outside of an accretion disk.

The more fundamental question is how such optically thick, warm slab of gas can be
created above an accretion disk of the lower temperature? 
What additional process heats up the warm corona and keeps it in the steady state with disk?

Many attempts were done to find what is the universal mechanism of energy dissipation in 
the warm corona, but the problem is still not fully solved.
Presence of strong corona which is responsible for most thermal energy release increases the stability of the disk \citep{1994-SvenssonZdziarski,1994-KusunoseMineshige,2007-BegelmanPringle-MagneticDisks}. 
However, detailed computations of radiative transfer in illuminated accretion disk atmospheres show that the outer warm/hot skin cannot be optically thick and stable at the same time \citep{ballantyne2001,nayakshin2001,2002-Rozanska-1,2004-Madej,2015-Rozanska}.
When irradiation increases, the ionized skin becomes unstable and most probably the gas is outflowing in the form of wind \citep{2015-Proga}.
Computations show that the warm/hot skin of the optical depth higher than 3 cannot be thermally stable (in pressure equillibrium) with cold accretion disk, when the skin is heated only radiatively \citep[][and references therein]{krolik1995,2000-Nayakshin,2002-Rozanska-1,2015-Rozanska}. 

\begin{figure}
    \resizebox{\hsize}{!}{\includegraphics{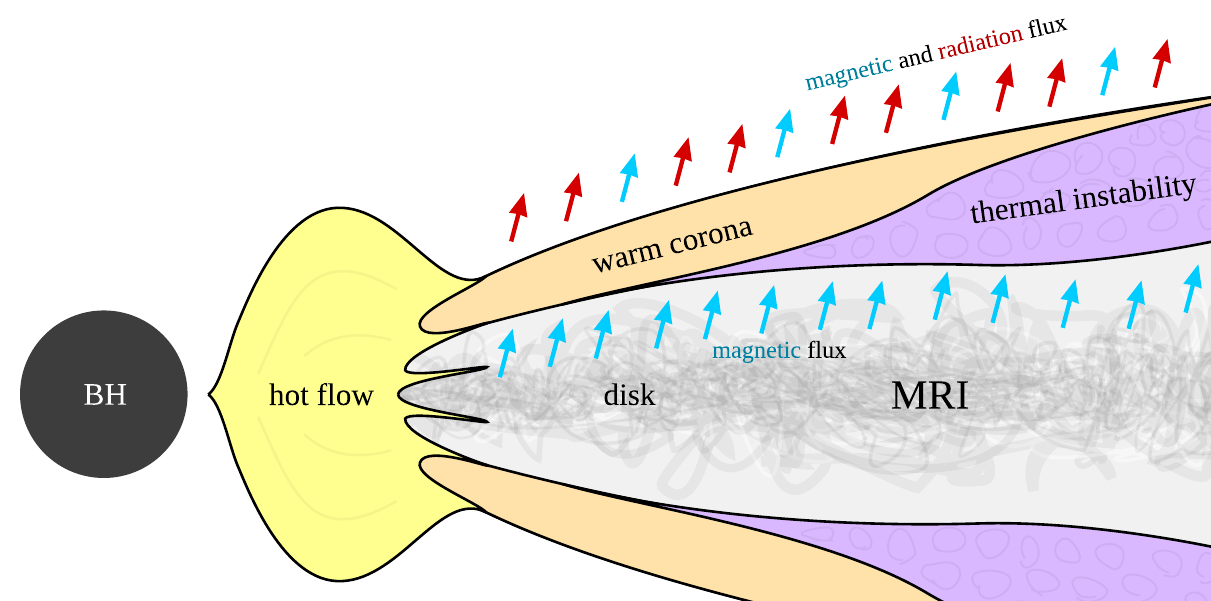}}
    \caption{Schematic illustration of a slice through the disk plane (only one
    	side was shown due to symmetry). Black hole is shown as the circle
    	on the left. Grey area represents optically thick and geometrically thin
    	disk when most of the accreting mass is located. Warm corona covering
    	the disk in which the magnetic energy is released as radiation is marked
    	in orange. Magnetic and radiative energy flux are represented by blue
    	and red arrows, respectively.
    	Part of the corona which possibly collapses due to thermal 
    	instability is shown in magenta. Inner optically thin and geometrically
    	thick hot flow, although not considered in our model, is drawn in yellow
    	for completion.}
    \label{fig:sketch}
\end{figure}

Additional heating of the warm corona layer by accretion process was also considered. 
However, in the fundamental paper of an accretion theory, \cite{1973-ShakuraSunyaev} introduced the geometrically thin disk model, where the kinetic energy of accreting gas is converted into thermal energy locally. Therefore, within the standard disk model all energy is dissipated deep inside the disk, at the equatorial plane. Vertical profiles of 
energy dissipation clearly show the exponential decrease towards the disk surface. 

The development of numerical simulations has shown the role of the magnetic field in accretion disk structure \citep{1991hawley,1991balbus,1992-ToutPringle,2001hawley,2008-Stone}. 
Many magnetohydrodynamic (MHD) simulations are available, but
none of them specifies the optical depth of the warm corona since most of them are
done without radiative cooling taken into account \citep[e.g.][]{2013-BaiStone,2013-Penna}. Those simulations, which contain simplified radiative cooling
\citep[e.g.][]{2004-Turner,2009-Hirose-RadiationDisks,2011-Ohsuga,2011-Noble} only discuss the existence of hot corona 
without constrains on its observational properties 
\citep[i.e.][]{2014-Jiang,2017sadowski}. Much advanced radiative transfer calculations 
usually are made a posteriori after general relativistic MHD simulations are completed \citep{2016-Schnittman}.
                                     
\citet[][hereafter RMB15]{2015-Rozanska} have developed an analytic model for an optically thick, uniformly heated, Compton cooled corona in the hydrostatic equilibrium. 
The heating mechanism was not specified in the model, but it allowed to integrate
vertically both disk and warm corona, coupled by mechanical heating and radiative cooling.
It allowed to determine the relations between the optical depth of the warm layer and the disk-corona energy budget.
Furthermore, the authors have shown that when some part of the gas pressure is 
replaced by magnetic pressure, the optical depth of stable corona increases. 

  In this paper, we assume the existence of geometrically thin and optically thick accretion
  disk around the black hole (see Fig.~\ref{fig:sketch}).
  We also assume that the accretion disk is magnetized and the magneto-rotational instability
  (MRI) is the primary source of viscosity and energy dissipation.
On these assumptions, we develop the model of slab-like warm corona covering the accretion disk,
adopting the  realistic heating of the gas by magnetic field reconnection which, contrary to the standard disk, 
increases towards the disk surface \citep{2006-Hirose}.  
We directly use the analytic formula derived  by \cite[][hereafter BAR15]{2015-Begelman}, where the vertical profile of magnetic heating of the accretion flow is determined.
On the top of this major assumption, the disk vertical structure together with the radiative transfer equation in gray atmosphere are fully solved with the relaxation method proposed by \cite{Henyey1964}. 
We have developed new numerical scheme which allows to solve non-linear differential equations in relatively short time, keeping the integration error low.  
Furthermore, with our new method we are able to pass through the thermally unstable 
regions which may arise whenever non-uniform heating and cooling mechanisms take place. 

As a result, we show for which range of parameters warm corona can exist in case 
of GBHB. We determine the optical depth of the warm layer, which is the main observable 
when analyzing X-ray data. The detailed vertical structure calculations allowed us 
to estimate an amount of energy dissipated in the corona in comparison to total 
energy released by accretion at a given radius. We show the radial structure of an optically thick corona
and give tight constraints for conditions for which such warm layer can exists. All results are 
compared with observations. 

Warm corona is not the only possible model to explain the soft X-ray excess.
Other models assume the relativistically smeared absorption \citep{2004-GierlinskiDone} or reflection from an accretion disk illuminated by hard X-ray continuum \citep{2019ApJ...871...88G} or bulk comptonization in central Compton cloud
\citep{2006ApJ...643.1098S,2016ApJ...821...23S}.
All of these models provide good fit to X-ray observations by incorporating additional Comptonized component, but they do not explain the physical mechanism that feeds energy into the warm gas to compensate for huge rate of Compton cooling.
Models that do provide the heating mechanism for a central, compact corona exist and can successfully explain some observations without the need for involvement of the magnetic field \citep{seifina2018}.
However, we note that the role of magnetic field in the accretion and jet formation process has been shown numerous times in the papers cited in this section and we do not consider its presence to be a strong assumption.
We show that accretion disk with operating magnetic dynamo is able to produce enough energy to heat up the extended surface layer
which has physical parameters that are consistent with these determined from spectral fitting to observations. Our results give
the possible physical explanation for the phenomenon which was in most cases only considered from observational perspective without analyzing the global energy budget. 
In order to verify whether our model can correctly reproduce spectral features seen in X-ray sources, spectral modeling
involving Compton scattering redistribution function must be performed.
That is, however, outside the scope of this paper.

The structure of the paper is as follows: 
Sec.~\ref{sec:mo} presents formalism of 
magnetically heated corona and disk/corona radiative transfer equation. 
It finishes with full description of differential equations and relaxation method used for their solution. 
Sec.~\ref{sec:res} presents results of our numerical computations. 
We display vertical structure of disk/corona system, but we also show radial limitation for which optically thick corona can exist. 
Discussion and conclusions are given in Sec.~\ref{sec:dis}.



\section{Set-up of the Model}
\label{sec:mo}



\subsection{Magnetically heated corona} 

\cite{2007-BegelmanPringle-MagneticDisks} and subsequently BAR15 have proposed a vertical model of magnetically supported disks (MSDs) driven by the magneto-rotational instability dynamo.
The MRI dynamo operating near the equatorial plane in the presence of external poloidal magnetic field, generates the toroidal magnetic flux, converting the mechanical energy of the accretion to the electromagnetic energy stored as tangled magnetic field lines. 
The total rate of this energy deposition per unit height is $\alpha_{\rm B} \Omega P_{\rm tot}$, where $\alpha_{\rm B}$ is the toroidal magnetic field production parameter, $\Omega$ is the Keplerian angular velocity and $P_{\rm tot}$ is the total pressure being a sum of gas, radiation and magnetic 
pressure.
Although the toroidal field production parameter $\alpha_{\rm B}$ does resemble the effective viscosity parameter $\alpha = t_{r \phi} / P$ (known from the classical thin disk model) in the sense that it scales the energy release within the disk, its physical meaning is different, even if expected values of both parameters are similar  \citep[Table 3, hereafter SSAB16]{2016-Salvesen}.

The magnetic flux ropes formed near the midplane rise buoyantly towards the surface, and in the model by BAR15, the velocity is very coarsely approximated by
\begin{equation}
 v_{\rm B}(z) = \eta \Omega z,
 \label{eq:vel}
\end{equation}
 where $0 < \eta < 1$ is a dimensionless parameter.
Similar process of buoyant emergence of magnetic fields towards the surface is observed on the Sun, however the dynamical properties and the origin of the magnetic field are vastly different.

During the motion of ropes upwards, the magnetic field decays by various processes, releasing its accumulated energy into heating the gas.
If we assume a purely toroidal flux rising vertically with speed $v_{\rm B}(z)$, the energy dissipation rate due to the induced current, $\mathcal{H}_{\rm mag}$, can be derived from the elemental electrodynamics and yields
\begin{equation}
\mathcal{H}_{\rm mag} = -v_{\rm B} dP_{\rm mag} / dz.
\label{eq:mag}
\end{equation}
This expression is consistent with intuitive reasoning: the energy loss is proportional to the magnetic energy density gradient times velocity.

Another process, which may heat the gas, is the magnetic reconnection due to cyclic reversals of toroidal field polarity observed in the simulations of the MRI dynamo.
The energy output of this process is proportional to the magnetic pressure $P_{\rm mag}$ and can be estimated as
\begin{equation}
\mathcal{H}_{\rm rec} = 2 \xi \Omega P_{\rm mag},
\label{eq:rec}
\end{equation}  
where $\xi$ is a proportionality constant (Appendix A in BAR15). 
The rate of thermal energy release peaks in the proximity of the disk photosphere ($\tau = 1$), however its exact distribution may change depending on the model parameters. 
Alongside $\xi$, we also define, after BAR15, the reconnection efficiency parameter $\nu$, which describes 
the ratio of reconnection process to the total magnetic viscosity: $ \nu = 2 \xi / \alpha_{\rm B} $, 
and it is very convenient for presentation of results (see Sec.~\ref{sec:poc}).

\begin{figure}
    \resizebox{\hsize}{!}{\includegraphics{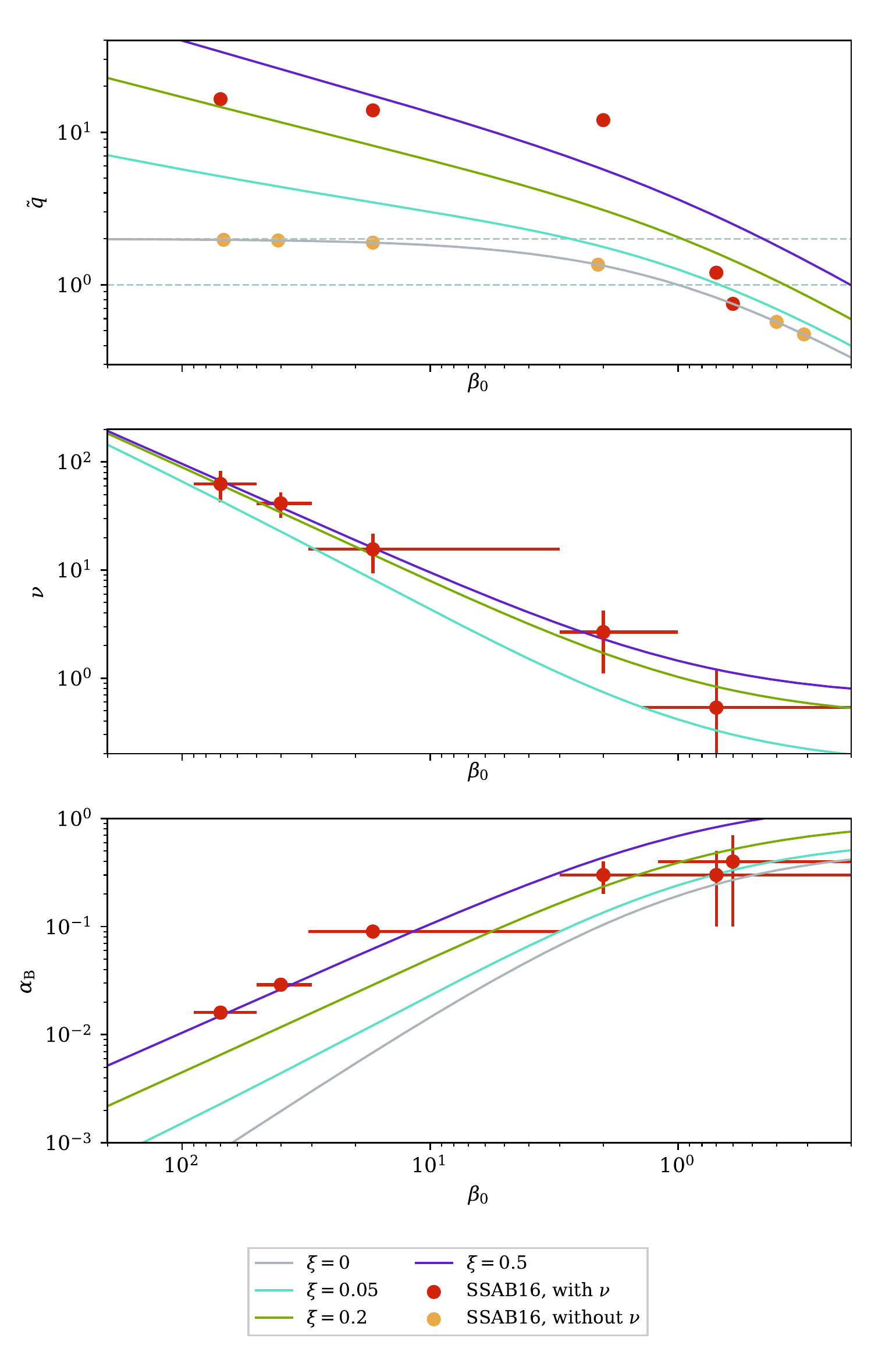}}
    \caption{
        Dependence of model parameters: toroidal field production efficiency - $\alpha_{\rm B}$ 
        (lower panel), reconnection efficiency - $\nu$ (middle panel), 
        and the magnetic field gradient parameter - $\tilde q$ (upper panel)
         on the magnetic parameter in the  disk midplane $\beta_0$. Points denote values 
         obtained from numerical simulations
        in SSAB16 and lines are our results from power law relation between $\alpha_{\rm B}$ 
        and $\eta$ (Eq.~\eqref{eq:etalfa}) for four values of $\xi$.
    }
    \label{fig:params}
\end{figure}

One consequence of such picture is that the field outflow eventually reaches the point where heating becomes inefficient and some portion of the accretion energy may be carried away by the Poynting flux and never released to heat the gas.
This is contrary to non-magnetic classical (neglecting advection) $\alpha$-viscosity models where all accretion power is released as the radiation.
The magnetic outflow may be associated with the matter outflow, but this association is not trivial: since the outflow speed $v_{\rm B}$ is less than the freefall speed $v_{\rm ff} = \Omega z$, most of the matter will probably slide down the magnetic flux tubes, only small amount being carried away by the outflow.
As a contrast, during the mass ejections in the solar corona, the timescale of the eruption is much shorter than the dynamical timescale, which results in much more efficient mass transport.

Obviously, numerical modeling shows that the process described above can be very complex, however this analytic approximation enables us to build and study the atmospheres of MSDs which is the purpose of this paper.

Following BAR15, the accretion energy is injected into the Poynting flux and consumed for gas heating, in varying proportion. If we express the Poynting flux as $F_{\rm mag} = 2 P_{\rm mag} v_{\rm B}$, the magnetic energy conservation equation can be written as
\begin{align}
\alpha_{\rm B} \Omega P_{\rm tot} 
&= \frac{dF_{\rm mag}}{dz} + \mathcal{H}_{\rm rec} + \mathcal{H}_{\rm mag} \nonumber \\ 
&= \frac{d}{dz} \left( 2 P_{\rm mag} v_{\rm B} \right) + \left(2 \xi \Omega P_{\rm mag} - v_{\rm B} \frac{dP_{\rm mag}}{dz} \right).
\end{align}
We substitute velocity and  different energy rates by Eqs.~\eqref{eq:vel}, \eqref{eq:mag}, and \eqref{eq:rec},
to obtain the final expression for the magnetic pressure gradient
\begin{equation}
\label{eq:magbil1}
 \eta \Omega z \frac{dP_{\rm mag}}{dz} =
\alpha_{\rm B} \Omega P_{\rm tot} - (2 \eta + \alpha_{\rm B}\nu ) \Omega P_{\rm mag}. 
\end{equation}
Finally, we can use the above equation to obtain the expression for the heating rate $\mathcal{H}$
[${\rm erg} \ {\rm cm}^{-3} \ {\rm s}^{-1}$] without using gradients
\begin{equation}\label{eq:heating}
\mathcal{H} = \mathcal{H}_{\rm rec} + \mathcal{H}_{\rm mag}
= 2 \left( \eta + \alpha_{\rm B} \nu \right) \Omega P_{\rm mag} - \alpha_{\rm B} \Omega P_{\rm tot}.
\end{equation}
The above heating rate describes the 
amount of released magnetic energy which heats the gas and then can be converted into radiation. 
In this paper we neglect advection, and assume that the system is fully cooled by radiation, which is specified in the next section.



\subsection{Radiative transfer in the disk/corona system}
\label{sec:tra}
We consider a gray, plane-parallel, optically thick atmosphere, which is additionally heated by dissipation of the magnetic field.
We include free-free process and Compton scattering in our model and neglect the synchrotron radiation which is not important in this regime of magnetic pressure considered (see Sec.~\ref{sec:regime} for exact numbers).
To determine the temperature structure, we consider three equations: a radiative equilibrium equation, transfer equation and Eddington approximation.

The radiative equilibrium equation for a small matter volume heated by various dissipation processes at rate $\mathcal{H}$, and by the local radiation field of mean intensity $J_\nu$ and cooled by 
the radiation emissivity $j_\nu$ reads
\begin{equation}\label{eq:radbil1}
4 \pi  \int_\nu \kappa_\nu \rho J_\nu d\nu + \mathcal{H} = 4 \pi j_{\rm bol},
\end{equation}
where $j_{\rm bol} \equiv \int_\nu j_\nu d\nu$, and $\rho$ is gas density in g\,cm$^{-3}$.
Here, we made an assumption that the entire dissipated thermal energy is locally converted into radiation (neglecting advection).

In an optically thick regime, when true absorption processes dominate over scattering, the radiation field is fully thermalized and strictly local, therefore for matter at temperature $T$, we have 
$J_\nu = B_\nu(T)$ where $B_\nu(T)$ is the Planck function for temperature $T$.
However, since in corona we are dealing with a strongly scattering medium this is not the case (particularly near the surface), and the actual radiation field intensity may deviate from Planck distribution towards Wien distribution.
The further deviation from the Planck law is caused by the presence of non-coherent inverse Compton scattering.
Despite the above, we assume that the spectrum of mean radiation intensity $J_\nu$ within the disk can be approximately described by  the Planck function with temperature $T_{\rm rad}$, i.e.: $J_\nu \approx B_\nu(T_{\rm rad})$. With these assumptions, we can define our frequency averaged quantities
\begin{align}
B &=  \int_{0}^{\infty}  B_\nu d\nu = \int_{0}^{\infty} B_\nu (T) d\nu = \sigma T^4 / \pi, 
\label{beee} \\
J &=  \int_{0}^{\infty}  J_\nu d\nu = \int_{0}^{\infty}  B_\nu (T_{\rm rad}) d\nu = \sigma T_{\rm rad}^4 / \pi.
\label{jeee}
\end{align}
Even if we assume that the radiation field has Planck distribution, we do not equal gas and 
radiation temperatures due to strongly scattering dominated corona.

For the gray atmosphere, considered in this paper, we  use both Rosseland and Planck averages, the latter denoted with ``P'' superscript.   
The following values for electron scattering and free-free opacities are assumed
\begin{align}
\kappa_{\rm es} &= 0.34 \ {\rm g}^{-1} {\rm cm}^{2}, \label{eq:kapsct}
\\
\kappa^{\rm P}_{\rm ff} &= 37 \cdot \kappa_{\rm ff} = 
\kappa_{\rm ff, 0}^{\rm P} \cdot \rho T^{-7/2} \ {\rm g}^{-1} {\rm cm}^{2}, \label{eq:kapabp}
\end{align}
where $\kappa_{\rm ff, 0}^{\rm P} = 37\cdot 6.21 \times 10^{22}$ is a constant from Kramers opacity approximation in cgs units.  Through this paper, we denote  total Rosseland mean opacity as $\kappa = \kappa_{\rm es} + \kappa_{\rm ff}$, while 
$\kappa^{\rm P} =  \kappa_{\rm es} + \kappa^{\rm P}_{\rm ff}$ is the total Planck opacity.


Using the gray approximation, we can rewrite the radiative equilibrium equation \eqref{eq:radbil1} as
\begin{equation}\label{eq:radbil2}
\mathcal{H} =  4 \pi \left( j_{\rm bol} -  \kappa^{\rm P} \rho J \right) \equiv \Lambda_{\rm rad} (\rho, T, T_{\rm rad}).
\end{equation}
We will refer to the function $\Lambda_{\rm rad} (\rho, T, T_{\rm rad})$ as the {\em net} radiative cooling rate.



To estimate the effect of inverse Compton scattering, we use standard formula given by \citep{RybickiLightman}, where by $J_{\rm C}$ we mean all incident photons that were scattered
with thermalized electrons of the gas temperature $T$:
\begin{equation}
 J_{\rm C}^\prime - J_{\rm C} =  J_{\rm C} \frac{ 4kT \langle \varepsilon \rangle - \langle \varepsilon^2 \rangle }{\langle \varepsilon \rangle m_e c^2 }.
\end{equation}
In this formula $J_{\rm C}^\prime - J_{\rm C}$ is the energy passed to the radiation from electrons by the inverse Compton scattering. The mean photon energy $\langle \varepsilon \rangle$ and mean squared energy $\langle \varepsilon^2 \rangle$ depend on photon energy distribution, and their ratio has following values
\begin{equation}
\langle \varepsilon^2 \rangle =
\begin{cases}
3.83 kT_{\rm rad} \cdot \langle \varepsilon \rangle & \text{for Planck spectrum} \\ 
4 kT_{\rm rad} \cdot \langle \varepsilon \rangle & \text{for Wien spectrum}.
\end{cases} 
\end{equation}

Although we assumed that radiation has Planck spectrum everywhere in the atmosphere, for a strongly scattering medium where the Compton cooled corona is present, the radiation spectrum is shifted towards slightly higher energies, and we find the approximation $\langle \varepsilon^2 \rangle = 4 kT_{\rm rad} \cdot \langle \varepsilon \rangle$ satisfactory for our model.
When the temperatures are low, density is high and free-free absorption dominates, this term is negligible anyway and does not change the result. Therefore, we replace $J_{\rm C} = \kappa_{\rm es} J$ to obtain the Compton term for the emission function:
\begin{equation}
j_{\rm IC} = \kappa_{\rm es} J \frac{ 4k(T- T_{\rm rad}) }{ m_e c^2 }.
\end{equation}

Finally, by taking into account all relevant opacities, we obtain the emission function $j_{\rm bol}$ of the following form
\begin{equation}\label{eq:sourcefun}
j_{\rm bol} = \kappa_{\rm ff}^{\rm P} B 
+ \kappa_{\rm es} J + j_{\rm IC} = \kappa_{\rm ff}^{\rm P} B 
+ \kappa_{\rm es} J \left[1 + \frac{ 4k (T - T_{\rm rad}) }{ m_e c^2 } \right].
\end{equation}
We can now substitute the expressions for $B$, $J$ and $j_{\rm bol}$  into the cooling function $\Lambda_{\rm rad}$ to obtain its final form
\begin{gather}
\Lambda_{\rm rad}  \left( \rho, T, T_{\rm rad} \right) \equiv  4 \sigma \rho  \left[  \kappa_{\rm ff}^{\rm P} \left( T^4 - T_{\rm rad}^4 \right)  + \kappa_{\rm es} T_{\rm rad}^4 \frac{ 4k\left( T - T_{\rm rad} \right) }{ m_e c^2 } \right] \nonumber
\\  
= 4 \sigma \rho \left( T - T_{\rm rad} \right) \left[  \kappa_{\rm ff}^{\rm P} \left(T + T_{\rm rad} \right) \left(T^2 + T^2_{\rm rad} \right)  + \kappa_{\rm es}  \frac{ 4kT^4_{\rm rad} }{ m_e c^2 }. \right] \label{eq:cooling}
\end{gather}
The net cooling rate can be split into two terms:
$\Lambda_{\rm rad} = \Lambda_{\rm B} + \Lambda_{\rm C}$, describing bremsstrahlung and Compton scattering contributions 
\begin{gather}
\Lambda_{\rm B} = 4 \sigma \rho  \kappa_{\rm ff}^{\rm P} \left( T^4 - T_{\rm rad}^4 \right) \ , \label{eq:coolingb}
\\
\Lambda_{\rm C} = 4 \sigma \rho \kappa_{\rm es} T_{\rm rad}^4 \frac{ 4k\left( T - T_{\rm rad} \right) }{ m_e c^2 } \ . \label{eq:coolingc}
\end{gather}

We solve here frequency integrated radiative transfer equation (Eq.~4 in RMB15), where zeroth 
moment of  this equation  reads
\begin{equation}
\frac{d H}{dz} = j_{\rm bol} - \kappa^{P} \rho J.
\end{equation}
If we average this over frequencies and replace the Eddington flux, with physical flux 
$4 \pi H_{\rm Edd} = F_{\rm rad}$, we obtain the expected result that the energy gained by the radiative flux is equal to the rate of radiative gas cooling
\begin{equation}\label{eq:rad2}
\frac{dF_{\rm rad}}{dz} =  \Lambda_{\rm rad}  \left( \rho, T, T_{\rm rad} \right) .
\end{equation}
Since we are interested in the optically thick regime, we assume the Eddington approximation 
 i.e. $J = 3K$, where $K$ is the second moment of specific intensity. 
We therefore solve the first moment of radiative transfer equation as
\begin{equation}
\frac{1}{3} \frac{d J}{dz} = \frac{d K}{dz} =  - \kappa \rho H_{\rm Edd}.
\end{equation}
Taking into account Eq.~\eqref{jeee} and the connection of Eddington flux with physical flux, 
the above equation yields
\begin{equation}\label{eq:rad3}
16 \sigma T_{\rm rad}^3\frac{d T_{\rm rad}}{dz} = -3 \kappa \rho F_{\rm rad}.
\end{equation}

Radiative transfer should be always coupled with the gas structure. We assume that the whole disk/corona system is in hydrostatic equilibrium i.e. 
\begin{equation}
 \frac{d P_{\rm gas}}{dz} + \frac{d P_{\rm mag}}{dz} = \frac{\kappa \rho}{c} F_{\rm rad} - \Omega^2 \rho z,
\label{eq:hydreq}
\end{equation}
where gas pressure typically is $P_{\rm gas}=k/{\mu m_H} T \rho$, and the first part on the right side denotes radiation pressure $P_{\rm rad}$. 
This equation is consistently implemented into full set of equations solved simultaneously from midplane to the surface of the corona. This treatment proves that we do not tread disk/corona system as two slab model, but we solve full gas structure together with proper heating processes and radiative cooling taken into account. 



\subsection{Equation set and boundary conditions}
\label{eq:boundary}

We solve the following set of five equations: \eqref{eq:magbil1}, \eqref{eq:radbil2}, \eqref{eq:rad2}, \eqref{eq:rad3}, and  \eqref{eq:hydreq}, using the expression for the heating rate given by Eq.~\eqref{eq:heating}, and for cooling function - by Eq~\eqref{eq:cooling}. The full numerical procedure of our numerical code
is presented in Appendix~\ref{appen1}. We integrate them  from the disk equatorial plane ($z = 0$) to the upper bound of the computational range ($z = z_{\rm max}$).
We estimate $z_{\rm max}$ by solving Eq.~38 in BAR15 for the height at which the magnetic pressure reaches $p=10^{-5}$
of the magnetic pressure at the midplane
\begin{equation}\label{eq:zmax}
z_{\rm max} = H_{\rm SS73} \sqrt{
\left( 4 + \frac{\alpha_{\rm B} \nu}{\eta} \right)
\left(p^{- 2 / \tilde q} - 1 \right)
},
\end{equation}
where $H_{\rm SS73}$ is the disk height derived from $\alpha$-disk model \citep{1973-ShakuraSunyaev} and
\begin{equation}\label{eq:qcor0}
\tilde q \equiv 2 + \frac{\alpha_{\rm B} (\nu - 1)}{\eta}.
\end{equation}

The requirement of hydrostatic equilibrium gives the balance between all pressures: magnetic, radiation and gas, and gravity force in vertical direction. 
We determine the gas temperature by solving the balance equation between the magnetic heating and Compton cooling in order to determine whether and in which circumstances the corona can form. 

We show, that the problem is better solved when other radiative processes, as free-free emission, are taken into account. 
We analyze two cases:  case ``A'', when $\Lambda_{\rm rad} \equiv \Lambda_{\rm B} + \Lambda_{\rm C}$ is given by the full Eq.~\eqref{eq:cooling} above,
   and case ``B'', when $\Lambda_{\rm rad} \equiv \Lambda_{\rm C}$ thus the free-free term is set to zero, and the atmosphere is only cooled by Compton scattering.

In the case ``B'', the Eq.~\eqref{eq:cooling}  is linear and we can get exactly one temperature solution in each point of the vertical structure, as opposed to case ``A'', when multiple solution can exist.
However, in case ``B'' the cooling rate is decreased, therefore in the transition region between the disk and corona, gas temperature is overestimated.
Despite that, since the density in the corona depends mostly on magnetic pressure gradient and becomes bounded to the gas pressure only in extremely weakly magnetized cases, the case ``B'' allows us to get almost exact density value compared to case ``A''.

We adopt the following procedure: we solve the entire model using case ``B''.
Now to correct the temperature in the transition region where case ``B'' performs the worst, we solve the Eq.\eqref{eq:rad2} again, but keep the density fixed, which guarantees to provide an unique solution.
Since we corrected the temperature but did not update the density or gas pressure, we have a small deviation from hydrostatic equilibrium.
However, the corona is dominated by the magnetic field pressure gradient, and this deviation is negligible and does not change the overall structure of the atmosphere.
At this little cost we are able to go past the instability and investigate the influence of disk magnetization on its presence and strength.



We adopt boundary conditions appropriate for accretion disk cylindrical geometry. 
The radiative flux carried through the midplane ($z = 0$) must vanish due to symmetry
\begin{equation}
F_{\rm rad} = 0.
\end{equation}

Following BAR15, we also assume that the magnetic pressure gradient at the equatorial plane must be zero.
Therefore, from Eq.~\ref{eq:magbil1} we get
\begin{equation}
P_{\rm tot} = \left( \frac{2\eta}{\alpha_{\rm B}} + \nu \right) P_{\rm mag}.
\end{equation}
If the radiation pressure is neglected, the relation derived by BAR15 between the parameters $\alpha_{\rm B}$, $\eta$, $\nu$ and the magnetic parameter ($\beta = P_{\rm gas} / P_{\rm mag}$) at the equatorial plane $\beta_0$ results:
\begin{equation}\label{eq:beta0}
\beta_0 = \frac{2\eta}{\alpha_{\rm B}} + \nu - 1.
\end{equation}

At the top of the atmosphere ($z \rightarrow \infty$), we assume that the sum of  the flux carried away by radiation and of the magnetic field is equal to the flux obtained by Keplerian disk theory
\begin{equation}
F_{\rm rad} +  F_{\rm mag} = F_{\rm acc} = \frac{3}{8\pi} \frac{GM\dot{M}}{R^3} \left(1 - \sqrt{\frac{3R_{\rm Schw}}{R}} \right),
\end{equation}
where  $G$ is gravitational constant, $M$ - black hole mass,
$\dot M$ - accretion rate, and $R$ is radial distance from black hole. Typically $R_{\rm Schw}$ denotes Schwarzschild radius given as $2GM/c^2$.  
We also use the standard boundary condition for radiative transfer, $J = 2H$, which guarantees that there is no external illumination of the disk.
Expressed in our convention, the boundary condition at $z_{\rm max}$ takes form
\begin{equation}
F_{\rm rad} - 2 \sigma T_{\rm rad}^4 = 0.
\end{equation}

This gives the total of four boundary conditions, which complete our set of five equations, since Equation~\eqref{eq:radbil2} is not a differential equation, and equations \eqref{eq:magbil1}, \eqref{eq:rad2}, \eqref{eq:rad3} and \eqref{eq:hydreq} are first-order, nonlinear, ordinary differential equations.

Our numerical computations are parametrized by six input quantities: three magnetic, $\alpha_{\rm B}$,
$\eta$ and $\nu$, and three disk parameters: black hole mass, radial distance and the accretion rate. 
For better convenience, we use accretion rate: $\dot m = \dot M \kappa_{\rm es} c / 48 \pi G M$, given
in units of Eddington rate with accretion efficiency for Newtonian potential.  

\begin{figure}
    \resizebox{\hsize}{!}{\includegraphics{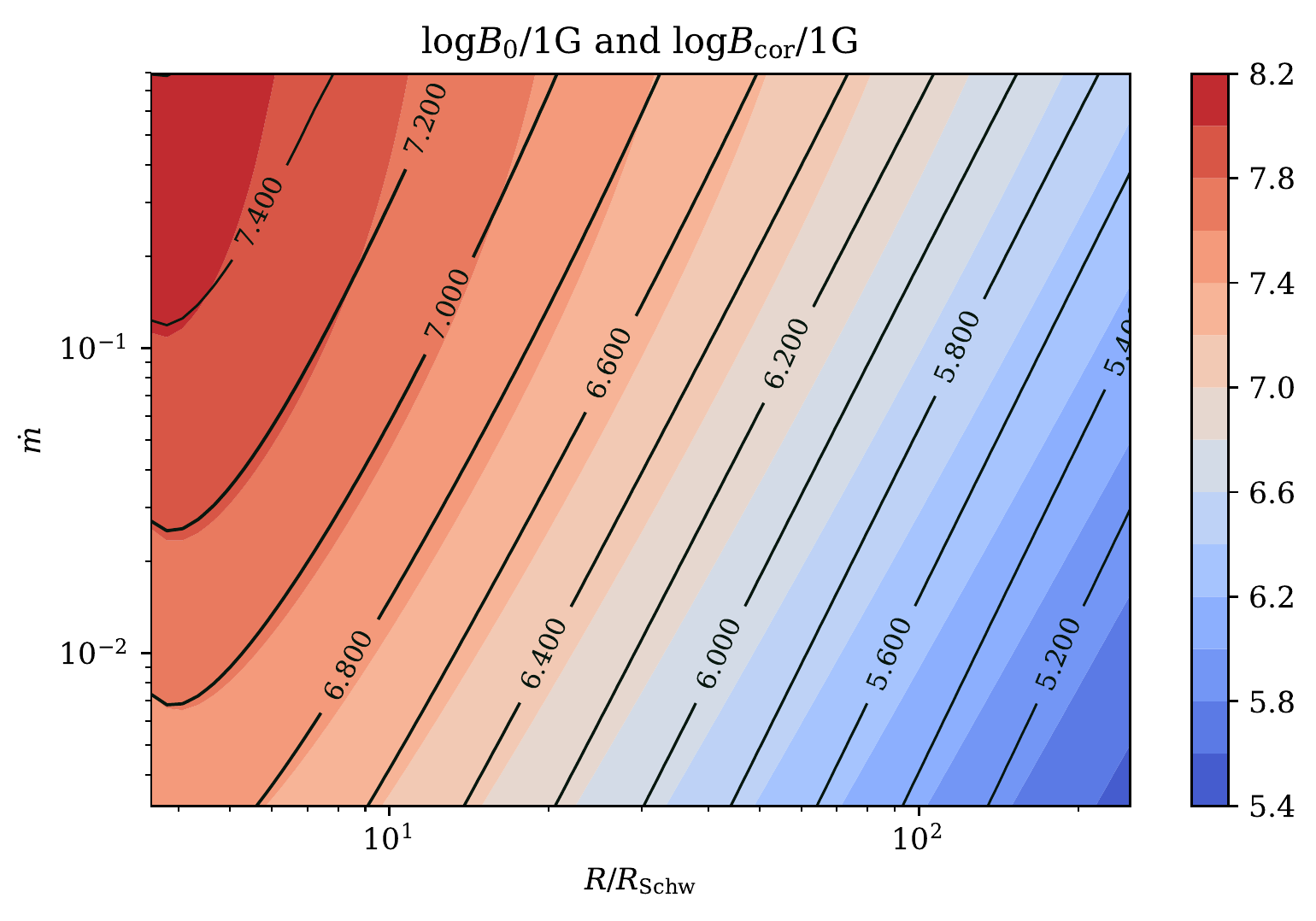}}
    \caption{Toroidal magnetic field at the equatorial plane $B_0$ and in 
        the corona $B_{\rm cor}$ at 
        optical depth $\tau_{\rm cor}$ (see text for definition) in Gauss units. Logarithmic values 
        of magnetic field are plotted on the radius ($R/R_{\rm schw}$) -- accretion rate ($\dot m$) parameter plane.}
    \label{fig:mag}
\end{figure}

\begin{figure*}
    \centering
    \includegraphics[width=18cm]{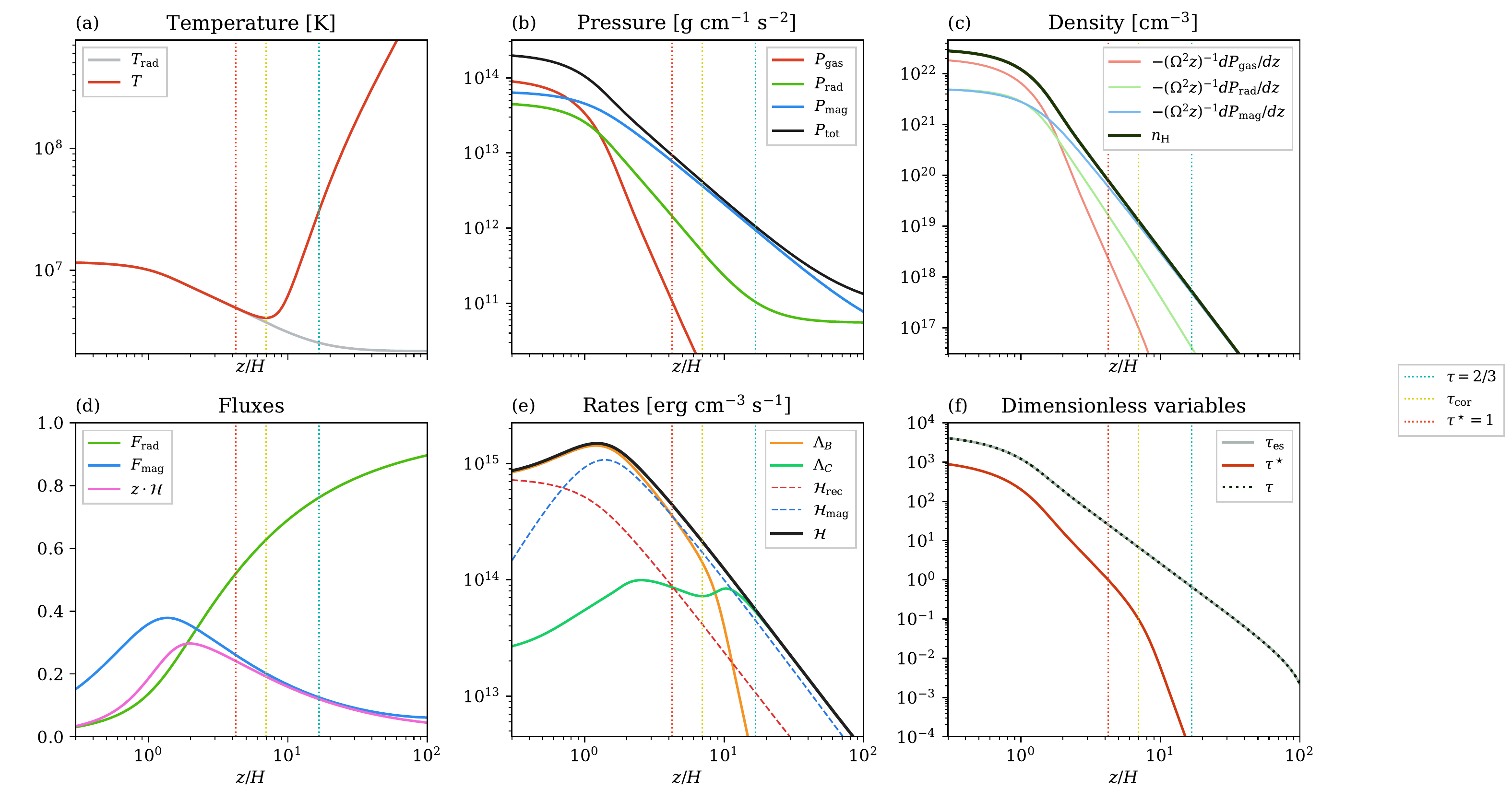}
    \caption{
        Vertical structure of the disk/corona system computed at  the single radius $R=10$~R$_{\rm Schw}$ for $\dot m=0.05$.   Horizontal axis is the the height above the 
     disk midplane given in scale height $H$ defined by Eq.~\eqref{eq:hdisk}.
        Three significant locations are marked with vertical dotted lines: photosphere (green), temperature minimum (dark yellow) and thermalization depth (red).
     Following panels display:
        a) radiation and gas temperature
        b) gas, radiation, magnetic and total pressure,
        c) gas density and pressure gradients,
        d) radiative energy flux, magnetic (Poynting) flux and gas heating rate
         all normalized to total accretion flux - $F_{\rm acc}$,
        e) net radiative cooling terms ($\Lambda_{\rm B}$ and $\Lambda_{\rm C}$) and gas magnetic heating rates,
        f) electron scattering, total and effective optical depths.
    }
    \label{fig:cute}
\end{figure*}

\begin{figure}
    \resizebox{\hsize}{!}{\includegraphics{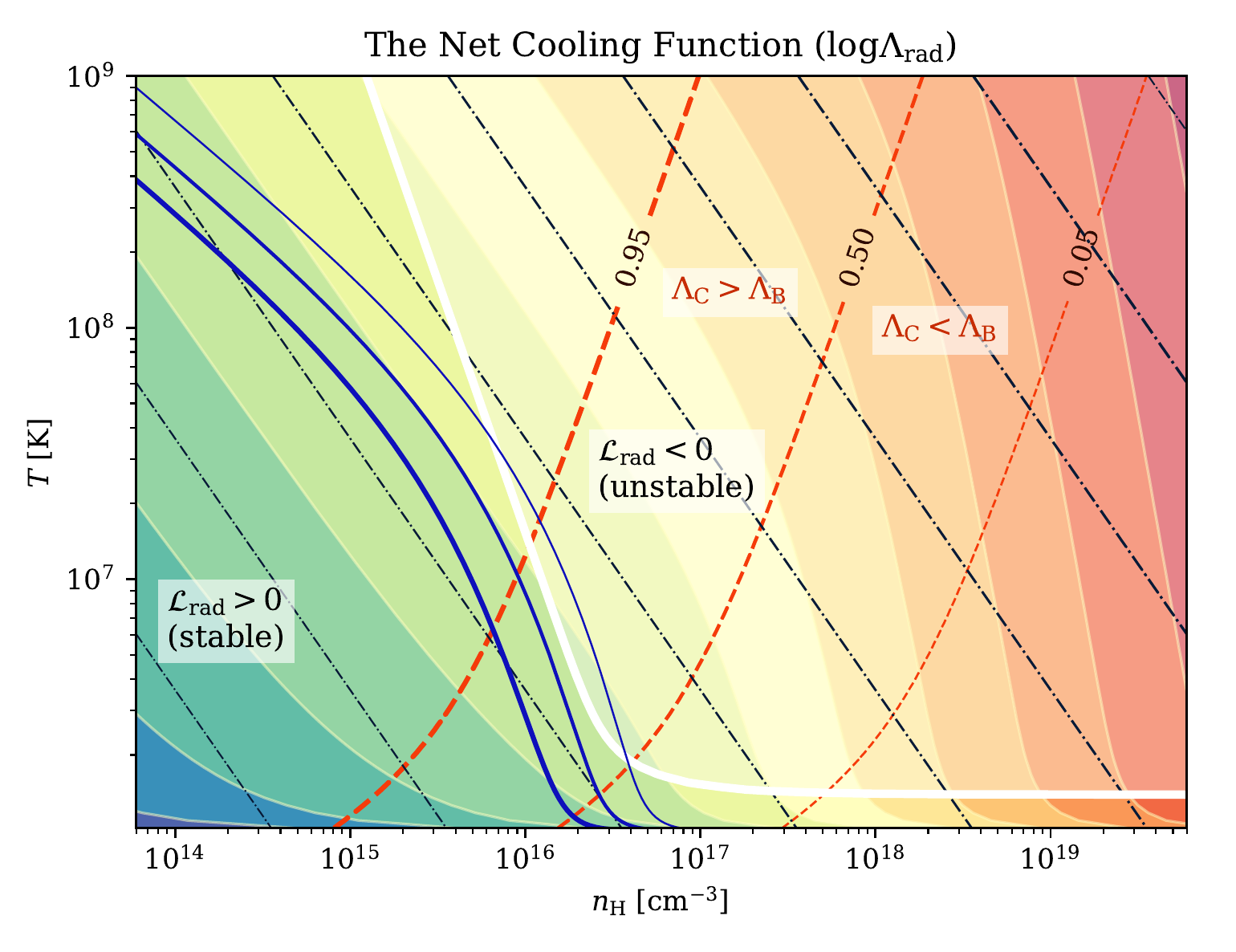}}
    \caption{
        Logarithm  of the net cooling rate $ \Lambda_{\rm rad}$ \,(Eq.~\ref{eq:cooling}), plotted on $n_{\rm H}$~--~$T$ parameter plane ranging from blue (low rate) to red (high rate). 
 The points of constant pressure are connected with black dash-dotted lines.
       The red dotted contours show the contribution of Compton cooling term to the total cooling.   The solid white line  shows where the cooling function is constant in isobaric perturbation, and separates the area of unstable solutions 
      where the  gradient of cooling function  $\mathcal{L}_{\rm rad} <0$.
        Dark blue lines of increasing width represent the vertical structure for 
        three accretion rates: 0.035, 0.050 and 0.072 respectively.
    }
    \label{fig:cooling}
\end{figure}

\begin{figure*}
    \centering
    \includegraphics[width=18cm]{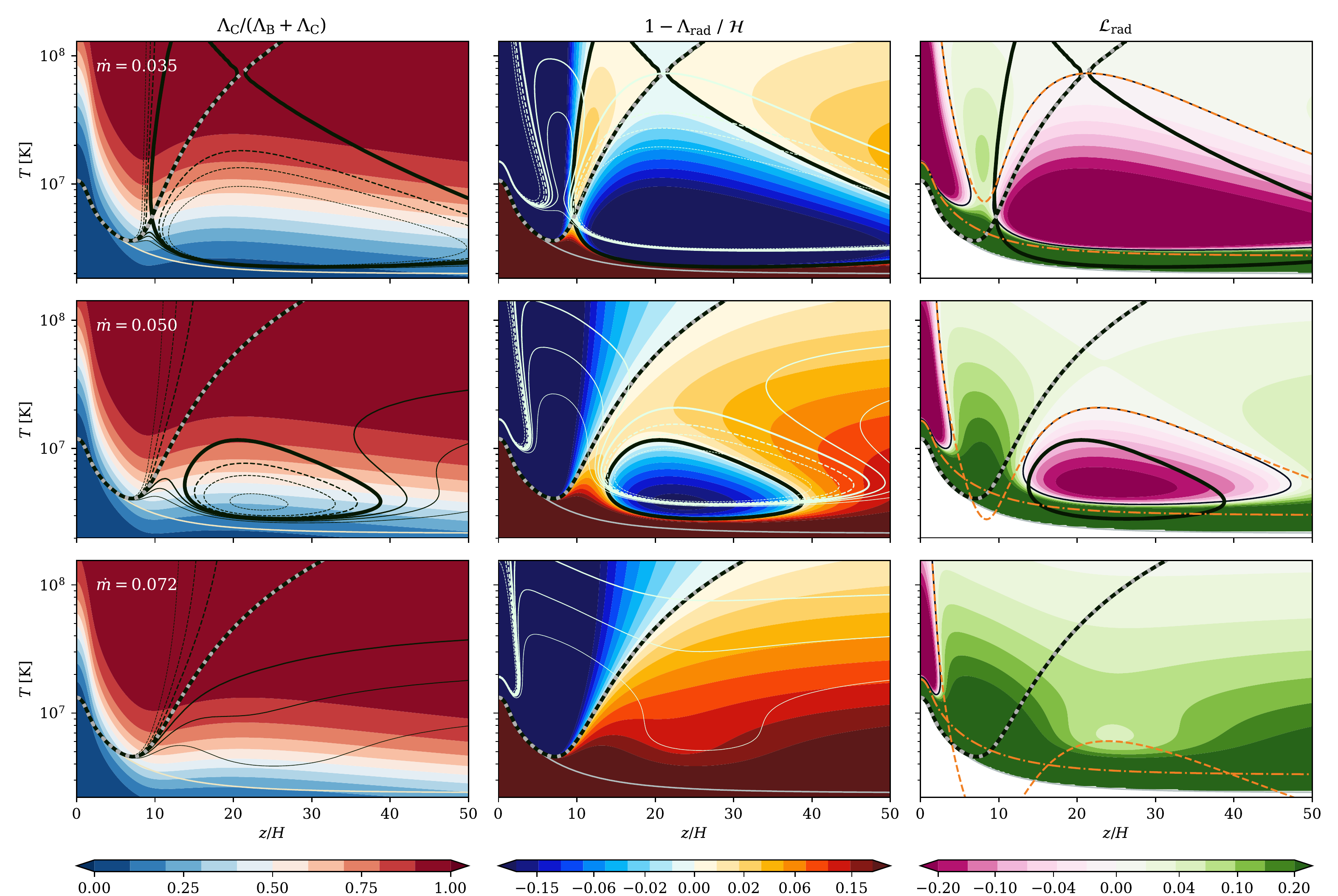}
    \caption{
        Onset of thermal instability in low temperature disk, $\dot m=0.035$ -- top panels, and its stabilization with the increase of the temperature, $\dot m=0.05$ and 0.072 -- middle and bottom panels.  Other disk parameters as given in  Sec.~\ref{sec:regime}. 
        At each panel the vertical axis represents the 
       possible gas temperature $T$ assuming constant gas pressure 
    ($\delta P_{\rm gas} = 0$), whereas the horizontal axis is the height above the disk midplane given in scale height $H$  (Eq.~\eqref{eq:hdisk}). 
   The possible solutions that satisfy the radiative energy balance are shown with thick black contours, while the solution obtained from our numerical computations is additionally marked with light gray dots. 
        A thinner gray line is used to show the radiation temperature, $T_{\rm rad}$.
       Color maps in the first panels column show the contribution from Compton scattering to the total cooling. In the second column,  the overall heating-cooling balance is shown: the yellow-orange areas are where the heating dominates cooling and blue areas is where cooling rate exceeds the heating.
        The third column presents the stability diagnostic --  the cooling function gradient $\mathcal{L}_{\rm rad}$ (Eq.~\ref{eq:instabil}), with green areas marking the stable regions and violet areas -- unstable regions. 
        Two limit solutions: \eqref{eq:instabilappx1} and \eqref{eq:instabilappx2} have been plotted with orange lines (dash-dotted and dashed, respectively) and show excellent agreement with numerical solution. Additionally, the content of neighboring plots (from right panels to the middle, and from middle to the left) have been superimposed as thin contours for the sake of clarity.}
    \label{fig:instabil}
\end{figure*}


\section{Vertical structure of the disk with corona}
\label{sec:res}


\subsection{Parameter regime}
\label{sec:regime}



Through this paper we consider the case of GBHB with $M=10$~M$_\odot$. The results are presented
for wide range of accretion rates from $\dot m = 10^{-3}$ up to 1 , and radial distances from 
marginally stable orbit up to $R/R_{\rm Schw} = 400$. The value of disk outer radius, adopted by us,
is taken from the fact that for larger distances corona does not exist at all or it is very weak. 

The parameters $\alpha_{\rm B}$, $\eta$ and $\nu$ used in our computations have been determined by
SSAB16 from numerical simulations for several runs of varying net poloidal magnetic flux value which resulted in different degree of disk magnetization.
Their results are plotted in Fig.~\ref{fig:params} by points.

For strongly magnetized case, simulations show that $\alpha_{\rm B} \approx \eta \simeq 0.3$, $\beta_0 \simeq 1$ and $\xi \simeq 0$ which gives the reconnection efficiency parameter $\nu = 0$. 
It means that the field production is efficient, the outflow is at high rate and dissipation by reconnection is small. On the other hand,
for weakly magnetized case, simulations result with $\alpha_{\rm B} \simeq 0.02$, 
$\eta \simeq 0.04$, $\beta_0 \simeq 50$, and  due to frequent dynamo reversals $\xi \simeq 0.5$. For such a case, the reconnection 
efficiency parameter $\nu \simeq 50$.
On the other hand, SSAB16 concluded that models with $\nu$ higher than a few tend to overestimate the magnetic field gradient for weakly magnetized models and $\nu = 0$ describes the magnetic pressure descent more accurately for these cases (see Fig.~13 from SSAB16).
This would be more consistent to assume $\nu$ parameter to have relatively small values, of the order of a few, to be 
in agreement with magnetic pressure vertical shape.

To reduce the number of magnetic parameters to consider, we assumed a constant value of $\xi$ and made a simplification that $\alpha_{\rm B}$ and $\eta$ are tied by a power law relation.
Using Eq.~\eqref{eq:beta0}, we recomputed $\alpha_{\rm B}$ which is consistent with assumed value of $\xi$ and $\eta^{\rm sim}$ and $\beta_0^{\rm sim}$ taken from Table~2 and 3 of the SSAB16 paper. 
\begin{equation}
\alpha_{\rm B} = \frac{\eta^{\rm sim} + \xi}{\beta_0^{\rm sim} + 1}
\end{equation}
We then performed  a power law fit and found that
for $\xi = 0.05$ the relation that best fits six data points is
\begin{equation}\label{eq:etalfa}
\eta = 1/3 \alpha_{\rm B}^{7/18} \ .
\end{equation}
The dependence of magnetic parameters with the use of above analytical formulae are plotted by lines 
in Fig.~\ref{fig:params} for comparison with simulations. From the figure 
we see, that we can adopt the values of magnetic parameters which are consistent with simulations.  

Upper panel of Fig.~\ref{fig:params}  shows that the magnetic pressure gradient which is 
an important quantity influencing the magnetic energy release. 
The magnetic pressure gradient can be derived from Eq.~\eqref{eq:magbil1} and yields
\begin{equation}\label{eq:qcor}
q = - \frac{d \ln P_{\rm mag}}{d \ln z} = 2 + \frac{\alpha_{\rm B} }{\eta} \left( \nu - \frac{P_{\rm tot}}{P_{\rm mag}} \right) \ .
\end{equation}
In the corona, when $P_{\rm gas} + P_{\rm rad} \ll P_{\rm mag}$, 
we can substitute $P_{\rm tot} / P_{\rm mag} \simeq 1$.
The value of $q$ is then constant and if we express it in terms of model parameters only, we get a result compatible with Eq.~37 in BAR15:
\begin{equation}\label{eq:xcor}
q \approx \tilde q =  2 + \frac{\alpha_{\rm B} \left( \nu - 1 \right)}{\eta} = \frac{2 \beta_0}{1 + \beta_0 - \nu} \ .
\end{equation}

With $\alpha_{\rm B}$ remaining the only free magnetic parameter in our model, the value that we find most suitable for potentially interesting disks is $\alpha_{\rm B} = 0.1$.
For such parameter, the toroidal magnetic field radial distribution at the equatorial plane $B_0$ 
and at the base of the corona $B_{\rm cor}$ (where the gas temperature achieves minimum), depending on an accretion rate and a distance from 
black hole is given in Fig.~\ref{fig:mag}. The magnetic field in our computations 
never approaches values of $10^8$~G.  


\subsection{Characteristics of the vertical profile}
\label{sec:vert}

Typical structure of the disk-corona system for $\dot m =0.05$ and $R/R_{\rm Schw} = 10$, is shown in Fig.~\ref{fig:cute}.
Results are presented with respect to the height above the disk midplane $z/H$, where 
half of the disk thickness is given by
\begin{equation}\label{eq:hdisk}
H = \sqrt{ \frac{\int_0^\infty \rho z^2 dz}{\int_0^\infty \rho dz}}.
\end{equation}
Furthermore, for better result presentation we define total optical depth of the system according to standard formula $ d\tau= -\kappa^{\rm P} \rho \, dz$,  electron scattering optical depth denoted as $ d\tau_{\rm es}= -\kappa_{\rm es} \rho \, dz$, and the density number per unit of volume as: $n_{\rm H} = \rho / m_H$.

Since in our model the transition in vertical structure between disk and warm corona occurs naturally, 
we reveal the location of the photosphere $\tau=2/3$ and thermalization depth $\tau^\star = 1$ where the seed photons originate \citep[see][Sec.~1.7]{RybickiLightman}, 
by vertical lines in Fig.~\ref{fig:cute}. The thermalization depth is equivalent to the 
effective optical depth, which in our case is computed using Planck opacities 
(see Sec.\,\ref{sec:tra}). We take the temperature minimum as the inner boundary of the corona, and  we refer to electron scattering optical depth of the temperature minimum as
$\tau_{\rm cor}$, which is also marked by vertical line at all panels of Fig.~\ref{fig:cute}. 


Despite that the geometrical proportions and physical parameters can vary when changing model input, we almost universally obtain structure where three distinct regions can be observed, which directly correspond to the gas pressure structure:
(i) accretion disk, where most of the matter is located, characterized by high density near the midplane, gas temperature exactly following the radiation temperature, and domination of the gas and radiation pressure over the magnetic pressure -- panel a) and b),
(ii) transition region, where bremsstrahlung is still an important cooling mechanism and gas temperature is coupled to radiation field temperature, but the magnetic pressure takes over in supporting the structure and slows the density decrease with height
	-- panel c)
(iii) corona, where the density is low and therefore the temperature must increase with height so the disk remains in  energy equilibrium; matter is cooled mainly by the inverse Compton scattering, while the density structure is fully supported by the magnetic pressure -- panel b) and c).
In the case presented in Fig.~\ref{fig:cute}, the following regions extend within 
$0 \leq z/H \lesssim 3$, $3 \lesssim z/H \lesssim 7$ and $7 \lesssim z/H$, respectively.
 Noticeable increase of the temperature appears in the transition region, when warm corona starts to form due to the magnetic heating. 

The overall magnetic pressure profile (see panel b) is in agreement with that obtained by 
BAR15, which confirms the correctness of our calculations. The density profile of the outermost coronal layers is fully shaped by pressure gradients, from which the magnetic pressure gradient is the largest (see panel c).

 When solving the vertical structure of the MSD, one needs to consider the 
 magnetic energy flux $F_{\rm mag}$,  in addition to radiative energy flux $F_{\rm rad}$,
  the latter being the  only energy transport channel in classical $\alpha$-disk.
Both fluxes normalized to $F_{\rm acc}$, are shown in Fig.~\ref{fig:cute}\,d), and are equal to zero in the disk midplane.
Contrary to the $\alpha$-disk, where the heating is mostly located near the disk midplane, we obtained the heating distribution which peaks at a few scale heights, although still below the temperature inversion point -- see Fig.~\ref{fig:cute}\,e). 
Such behavior is mostly the effect of purely magnetic heating $\mathcal{H}_{\rm mag}$, which depends on the magnetic pressure gradient.
Heating by reconnection $\mathcal{H}_{\rm rec}$ is proportional to the magnetic pressure (rather than its gradient) and peaks around the midplane.

The total optical depth of the disk/corona system is of the order of $10^4$.
The differences between various optical depth profiles are  visible in 
panel f) of Fig.~\ref{fig:cute}. For the above example, the whole corona is 
dominated by electron scattering i.e.: $\tau_{\rm es}=\tau$. The corona starts at
 $\tau_{\rm cor} \sim 10$,
and total optical depth $\tau \sim 20$ where full thermalization occurs e.g. 
$\tau^\star = 1$. The corona is optically thick being in agreement with recent observations.


\begin{figure*}
    \centering
    \includegraphics[width=18cm]{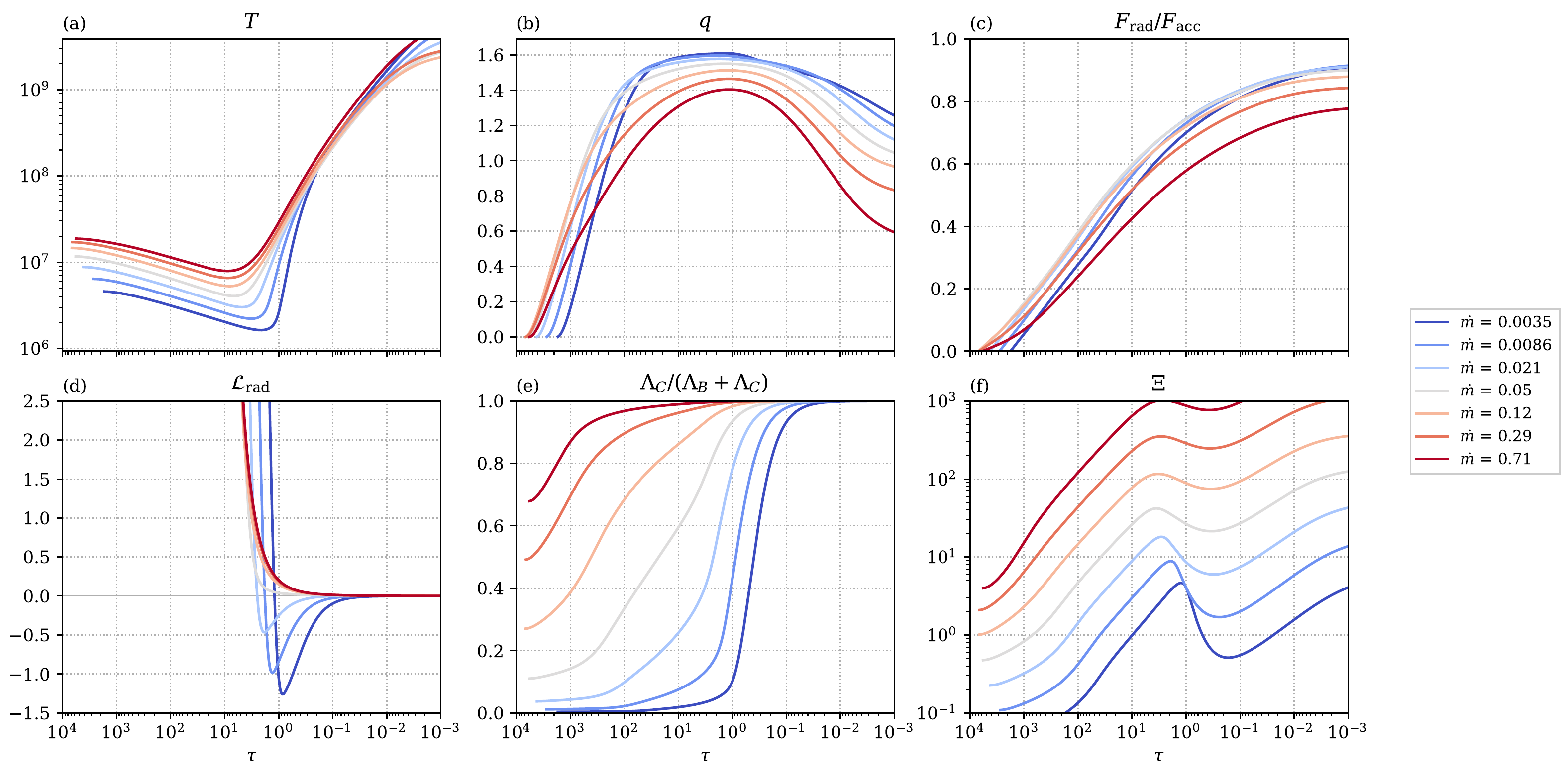}
    \caption{
       Vertical structure of the  disk/corona models computed at $R=10$~R$_{\rm Schw}$, for different values of the accretion rate ($\dot m$), plotted against the total optical depth $\tau$.
        Panels show:
        (a) gas temperature,
        (b) gradient of magnetic pressure (Eq.~\ref{eq:qcor}),
        (c) radiative energy flux normalized to total accretion flux - $F_{\rm acc}$, 
        (d) the cooling function gradient (Eq.~\ref{eq:instabil}),
        (e) contribution of Compton cooling to total cooling,
        (f) ionization parameter $\Xi $ (Eq.~\ref{eq:ionxi}).
    }
    \label{fig:multi-M-tau}
\end{figure*}

\begin{figure*}
    \centering
    \includegraphics[width=18cm]{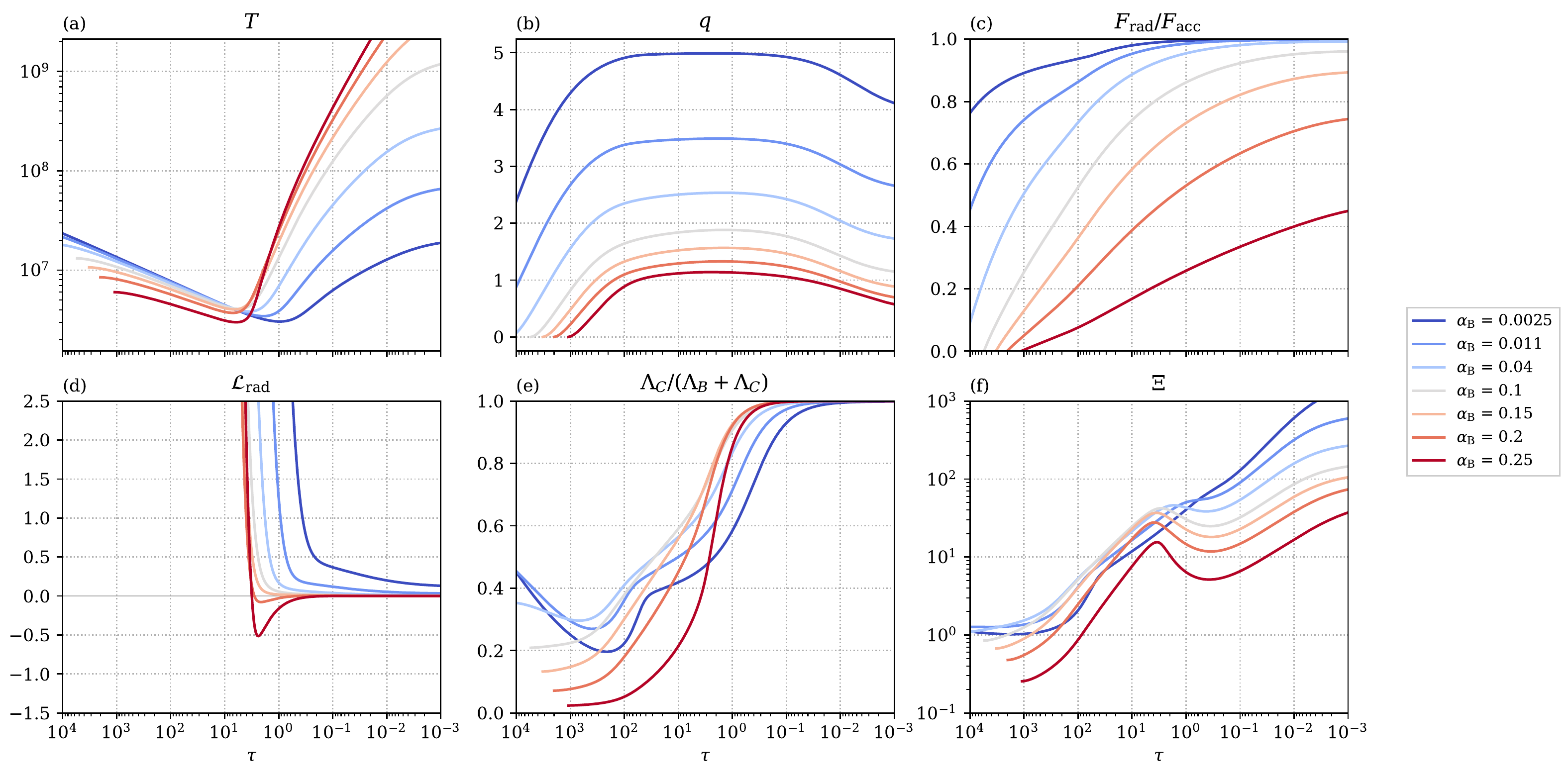}
    \caption{Same as in Fig.~\ref{fig:multi-M-tau}, but for 
    different values of magnetic parameter $\alpha_{\rm B}$.}
    \label{fig:multi-A-tau}
\end{figure*}


\subsection{Thermal instability of the intermediate layer}

\label{sec:instabil}

Thermal instability in the medium can occur if under given conditions, cooling rate does not increase with temperature.
Thermal balance equation has then three solutions, and if the matter is on the unstable branch, a thermal runaway or collapse occurs following a small deviation from the equilibrium.
The existence of coronal thermal instability in accretion disks has been investigated for 
many years \citep{krolik1981,1994-KusunoseMineshige,1993-NakamuraOsaki,1996-RozanskaCzerny,1999-Rozanska}, nevertheless it was never treated in the presence of magnetic field. 

Considering the static vertical structure of the accretion disk, it is reasonable to assume that any thermal collapse is an isobaric process.
Condition for thermal instability will read
\begin{equation}\label{eq:instabil}
\mathcal{L}_{\rm rad} \equiv \left. \frac{d \ln \Lambda_{\rm rad}}{d \ln T} \right|_{\delta P_{\rm gas} = 0}  < 0 \ .
\end{equation}
If we split the cooling into $\Lambda_{\rm B}$ and $\Lambda_{\rm C}$ contributions and substitute opacities given by Eq.\,\eqref{eq:kapsct} and \eqref{eq:kapabp}, we obtain
\begin{align}
\left. \frac{d \Lambda_{\rm B}}{dT} \right|_{\delta P_{\rm gas} = 0}  &= 2 \sigma \frac{\kappa_{\rm ff}^{\rm P}}{T} \left( 11 T_{\rm rad}^4 - 3 T^4 \right) \ ,
\\
\left. \frac{d \Lambda_{\rm C}}{dT} \right|_{\delta P_{\rm gas} = 0}  &= 16 \sigma \frac{\kappa_{\rm es} \rho}{T} \frac{kT_{\rm rad}^5}{m_e c^2} \ .
\label{eq:divcomp}
\end{align}

Depending on the contributions from the above two terms, we distinguish tree cases: 
i) when $\Lambda_{\rm B} \ll \Lambda_{\rm C}$, the solution is always thermally stable, since Eq.\,\ref{eq:divcomp} has positive sign,
 ii) when  $\Lambda_{\rm B} \gg \Lambda_{\rm C}$, instability occurs for
    \begin{equation}\label{eq:instabilappx1}
    T \ge \sqrt[4]{\frac{11}{3}} T_{\rm rad} \approx 1.38 T_{\rm rad},
    \end{equation}
and finally, iii) when $\Lambda_{\rm B} \approx \Lambda_{\rm C}$ and $T > T_{\rm rad}$, radiative heating by bremsstrahlung is negligible, and the criterion for instability reads
    \begin{equation}\label{eq:instabilappx2}
    \rho^2 T \ge \left[ \frac{8}{3} \frac{\kappa_{\rm es}}{\kappa_{\rm ff,0}^{\rm P}} \frac{kT_{\rm rad}^5}{m_e c^2} \right]^2 \approx 4.30 \cdot 10^{-9} \left( \frac{T_{\rm rad}}{10^6 \ \mathrm{K}} \right)^{10}.
    \end{equation}
 In isobaric regime, when $\delta P_{\rm gas} = 0$ or $\rho T = {\rm const.}$, this defines upper limit for temperature for which thermal instability occurs. 
The transition between these regimes ii) and iii) occurs for the density
    \begin{equation}\label{eq:instabilappx3}
    \rho = \frac{8}{3} \sqrt[8]{\frac{3}{11}} \frac{\kappa_{\rm es}}{\kappa_{\rm ff, 0}^{\rm P}} \frac{kT_{\rm rad}^{9/2}}{m_e c^2}
    \approx 5.58 \cdot 10^{-35} T_{\rm rad}^{9/2}.
    \end{equation}


In order to show the transition between stable and unstable corona, 
the net cooling function in MSD for three slightly 
varying accretion rates: 0.35, 0.50 and 0.72 is presented in Fig.~\ref{fig:cooling}
for the radiation temperature $T_{\rm rad} = 10^6 {\rm K}$. 
Three models vary mainly by the shift in density, the highest density corresponding to the lowest accretion rate.

Each model starts at the disk midplane where $T \approx T_{\rm rad}$, and only when the transition to corona starts the gas temperature diverges from radiation temperature.
If the density is higher than that given by Eq.~\ref{eq:instabilappx3}, the model will inevitably enter the zone defined by Eq.~\ref{eq:instabilappx1}.
Otherwise, for hotter disks, the Compton cooling contribution will yield at least half
of the total cooling, and model might graze the instability boundary defined by Eq.~\ref{eq:instabilappx2} but never go past it.
This means that every but the Compton dominated disks will be affected by thermal instability of the transition region. The only stable solution is possible for
relatively rare disks for the low cooling rate i.e. left-hand side of white thick solid line. 

Same three models are shown Fig.~\ref{fig:instabil}.
In order to identify all thermal equilibrium solutions for each height above the disk midplane, Compton cooling fraction, energy balance and cooling function gradient ${\cal L}_{\rm rad}$ have been shown depending on gas temperature $T$ while maintaining the constant gas pressure.
For the hottest model (highest accretion rate), there is unambiguous solution in each point.
This changes when the temperature decreases: although the main, continuous solution is still stable (compared to Fig.~\ref{fig:cooling}), two cooler solutions appear.
The middle solution lies within unstable zone (pink area in third column), but the coldest solution is stable.
It means that since this area is magnetic pressure dominated, thus abrupt density change would not affect the hydrostatic equilibrium very much and the existence of two-phase matter is possible.
Finally, for the lowest accretion rate (upper panels of Fig.~\ref{fig:instabil}), no numerically consistent stable solution exists.
We suspect that matter in this region would remain cold, but above unstable region the corona might still form.
This corona would be hot and optically thin, and in discussed example the transition would occur at about $z / H \sim 20$.

Equations \eqref{eq:instabilappx1} and \eqref{eq:instabilappx2} define the lower and upper limit for temperature that allows the occurrence of thermal instability.
This can be clearly seen in Fig.~\ref{fig:cooling}, as each isobaric contour slices the instability boundary in exactly two points.
These solutions have also been plotted in Fig.~\ref{fig:instabil} with orange lines.
One can see that they indeed suffice to describe the instability in most of the vertical structure of the disk.

\subsection{Local properties of the corona} 
\label{sec:loc}

The disk/corona transition at the given radius  depends on the value of local temperature 
and density. We present here the evolution of the selected disk parameters with an accretion rate and magnetization state in the aim to 
demonstrate better how the  warm corona can arise at radius $R=10$~R$_{\rm Schw}$ . 
For this purpose we define here two additional quantities related to the radiation pressure, since 
the radiation energy is the main observable of the accretion disk/corona systems. 
The first such quantity is so called radiation efficiency i.e. the ratio of the radiation flux to the total flux dissipated locally at the certain 
point of vertical structure due to magnetic field reconnection $F_{\rm rad}/F_{\rm acc}$, while the second 
quantity is an ionization parameter $\Xi$ \citep{krolik1981} defined as  
\begin{equation}\label{eq:ionxi}
\Xi = \frac{P_{\rm rad}}{P_{\rm gas}} \ .
\end{equation}

Thus, at Figs.~\ref{fig:multi-M-tau} and \ref{fig:multi-A-tau} we show the structure of 
temperature, magnetic field gradient,  radiation efficiency, cooling function gradient, 
contribution of Compton cooling to the total cooling and ionization parameter 
at panels a), b), c), d), e), and f) respectively. 
For all cases of the accretion rate we achieve  hot outermost layer on the top of the inner accretion 
disk.

The temperature of corona  is very stable with respect to accretion rate, nevertheless
the optical depth for which the disk/corona transition occurs increases with accretion rate
reaching the value of $\approx 10$ for $\dot m =0.71$.
In contrast, the accretion rate determines the density of the disk and consequently the value of magnetic pressure gradient which saturates in the disk at the highest value for the lowest accretion. 
At the point of disk/corona transition magnetic pressure slightly decreases, but the rate 
the rate of radiative energy release is roughly the same, see panel c).
We conclude that the primary factor causing the optical depth of the corona to vary are the changes in disk temperature (and the transition region, consequently).

Panel~e) partly explains this behavior: for lower accretion rates, cooling in the disk is dominated by bremsstrahlung, which switches off at around $\tau \sim 1$, causing an abrupt temperature gradient in corona boundary. At the same time, cooling function gradient also becomes negative for those cases (panel d).
We conclude here, that this contribution from bremsstrahlung also causes the formation of unstable region, as discussed in Sec.~\ref{sec:instabil}.
For higher accretion rates, the contribution from bremsstrahlung is not that significant, and transition from disk to the corona is almost seamless. The ionization parameter presented in 
panel f) does not indicate thermally unstable region. It shows the inversion with optical depth which is shallower for Compton dominated cases (high accretion rate). And finally, $\Xi$ is larger 
for hotter and more rare disks.  

The changes in corona are much more dramatic when dependence on magnetic field is examined, as seen in Fig.~\ref{fig:multi-A-tau}.
Panel a) shows that corona optical depth and temperature are radically different for weak and strong magnetic field cases.
For the weak field, disk is much hotter and radiation flux almost saturates to its maximum value well inside the disk (panel c) blue line), in contrast to strong field case, where disk is much colder, 
 and more than half of the thermal energy is released in the corona above $\tau = 1$ (panel c) red line). For the corona the temperature behavior is exactly opposite. 
 
In case of dominant role of the magnetic pressure, 
in the corona the magnetic pressure changes as $P_{\rm mag} \propto z^{-\tilde q}$, density as $\rho \propto z^{-(\tilde q+1)}$ and heating rate as $\mathcal{H} \propto z^{-\tilde q}$.
High values of $q$ mean very steep decline of magnetic field strength and gas heating rate, which in turn means that most energy will be dissipated within the disk, and corona will be weak.  
This behavior is nicely demonstrated at panel b) of Fig.~\ref{fig:multi-A-tau}.
Stronger magnetic field yields to the effective disk cooling and consequently results in the appearance
of thermal instability in disk/corona transition (panel d). 
Despite large temperature differences, $\Xi$ within the disk remains almost constant (panel e), while in the corona it decreases with magnetization even if coronal temperature is higher.


\subsection{Global properties of the corona}
\label{sec:poc}

\begin{figure*}
    \centering
    \includegraphics[width=18cm]{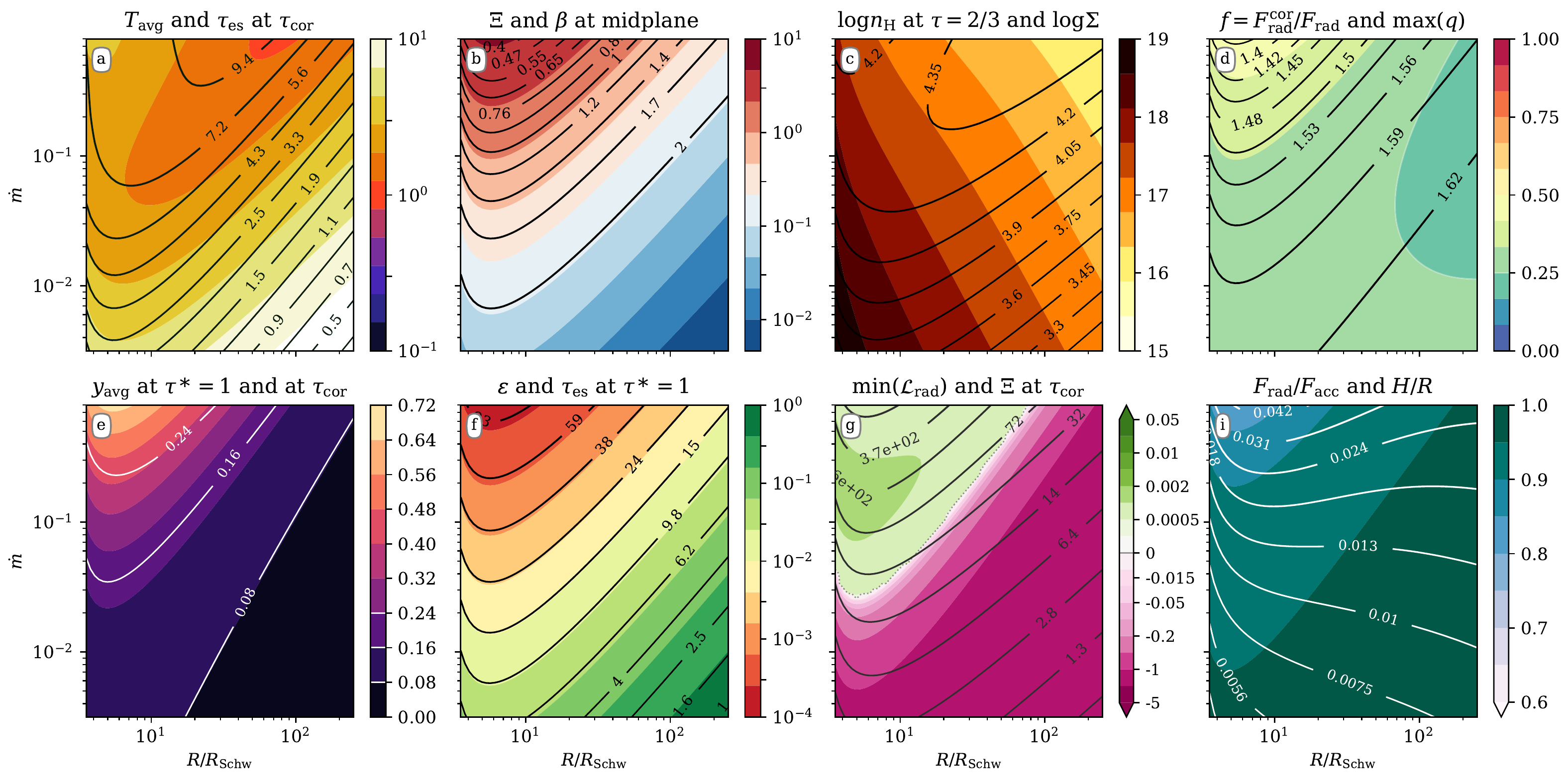}
    \caption{Properties of the corona on the radius ($R / R_{\rm schw}$) -- accretion rate ($\dot m$) parameter plane. 
        All models have been calculated for $M_{\rm BH} = 10 M_\odot$, $\alpha_{\rm B} = 0.1$,  
        and $\xi = 0.05$. Panels display:
        (a) average temperature  of the corona (Eq.~\ref{eq:tavg}) in keV (colors) and $\tau_{\rm cor}$ (contours);
        (b) ionization parameter $\Xi$ (colors) and magnetic $\beta$ parameter (contours)
        both at the disk midplane;
        (c) number density at $\tau=2/3$ (colors) and total column density $\Sigma$ in g~cm$^2$ (contours);
        (d) fraction of radiative energy produced by the corona $f$ (Eq.~\ref{eq:chipar}, colors),
        the maximum value of magnetic field gradient $q$ (Eq.~\ref{eq:xcor}, contours);
        (e) $y$ parameter of the warm corona (Eq.~\ref{eq:yavg}), computed up to thermalization layer $\tau^\star = 1$ (colors) and to $\tau_{\rm cor}$ (contours);
        (f) photon destruction probability $\epsilon = \kappa_{\rm ff}^{\rm P} / \kappa^{\rm P}$ (colors) and electron scattering optical depth ($\tau_{\rm es}$) (contours) at the thermalization layer;
        (g) the minimum value of $\mathcal{L}_{\rm rad}$ (Eq.~\ref{eq:instabil}) throughout the disk height (colors), green areas are stable disks, whereas magenta areas indicate thermal instability in the corona, the gray dotted line corresponds to $\mathcal{L}_{\rm rad} = 0$, and 
        ionization parameter $\Xi$ at corona base (contours);
        (h) radiative energy flux normalized to total accretion flux (colors),
         and disk scale ratio $H / R$ (contours).
    }
    \label{fig:maps-MR}
\end{figure*}

\begin{figure*}
    \centering
    \includegraphics[width=18cm]{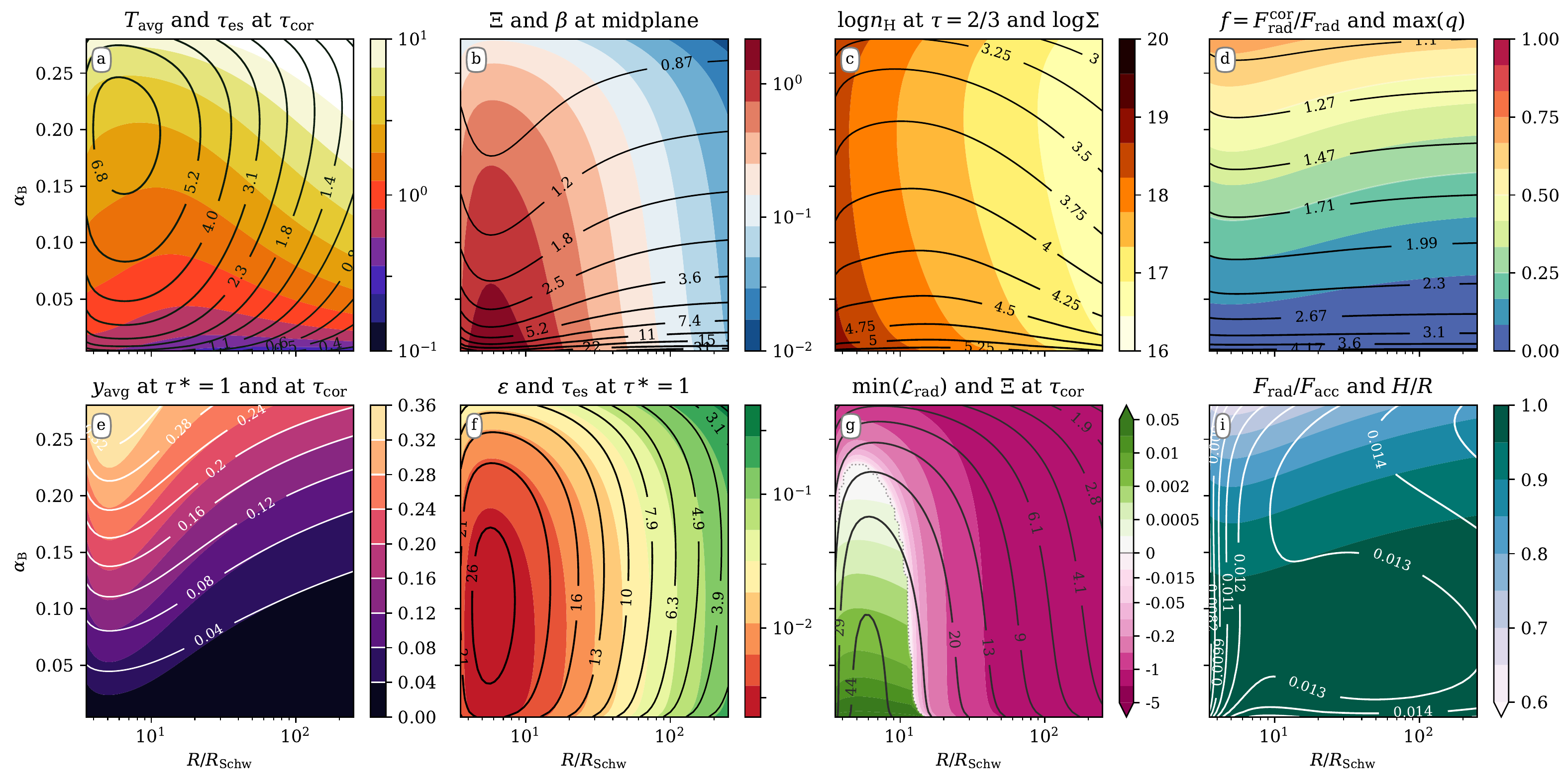}
    \caption{Same as in Fig.~\ref{fig:maps-MR} but on  the radius ($R / R_{\rm schw}$) -- toroidal field production parameter ($\alpha_{\rm B}$) plane.}
    \label{fig:maps-AR}
\end{figure*}

\begin{figure*}
    \centering
    \includegraphics[width=18cm]{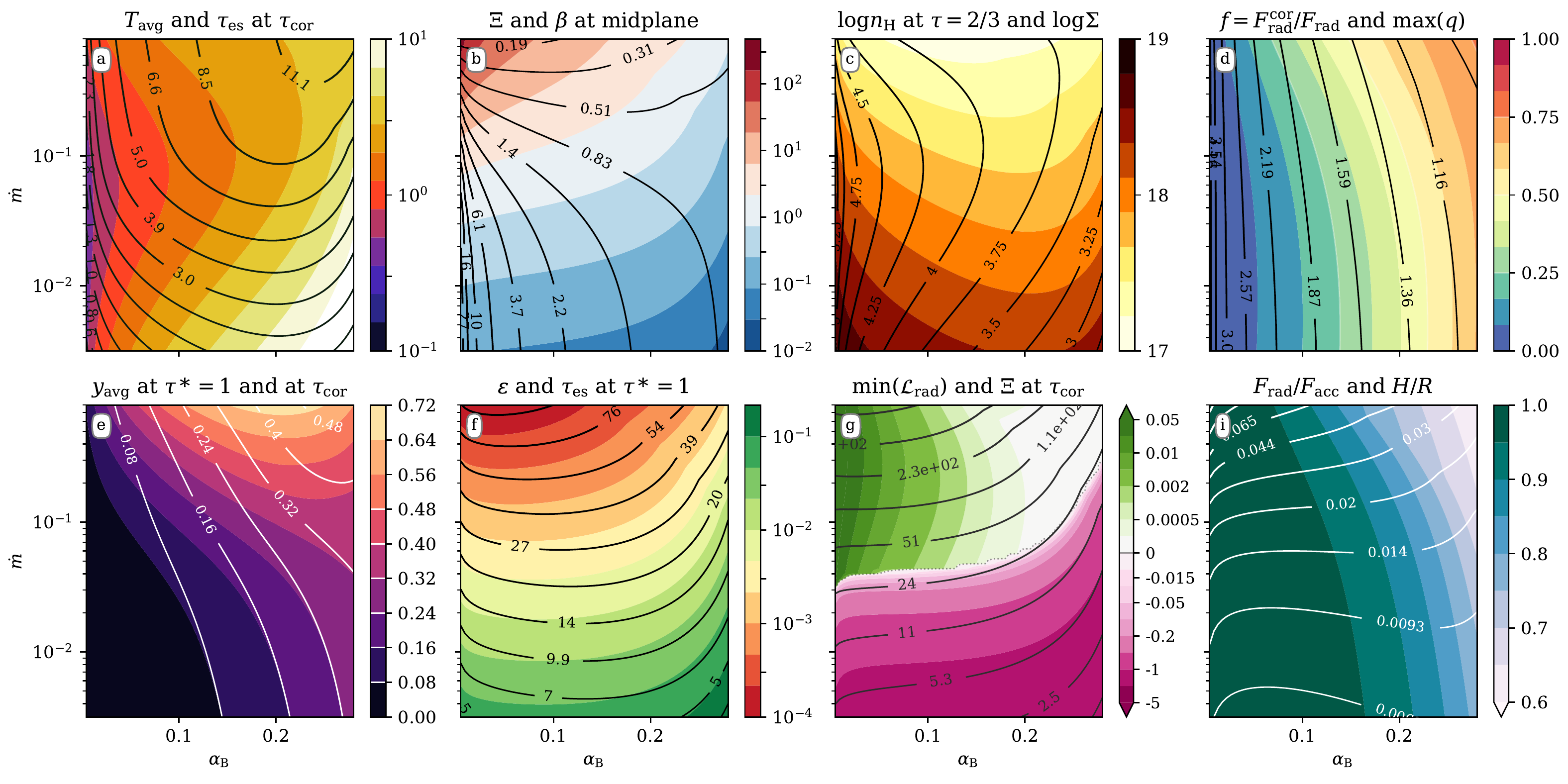}
    \caption{Same as Fig.~\ref{fig:maps-MR} but on  toroidal field production parameter ($\alpha_{\rm B}$) -- accretion rate ($\dot m$) plane.}
    \label{fig:maps-MA}
\end{figure*}

\begin{figure*}
    \centering
    \includegraphics[width=18cm]{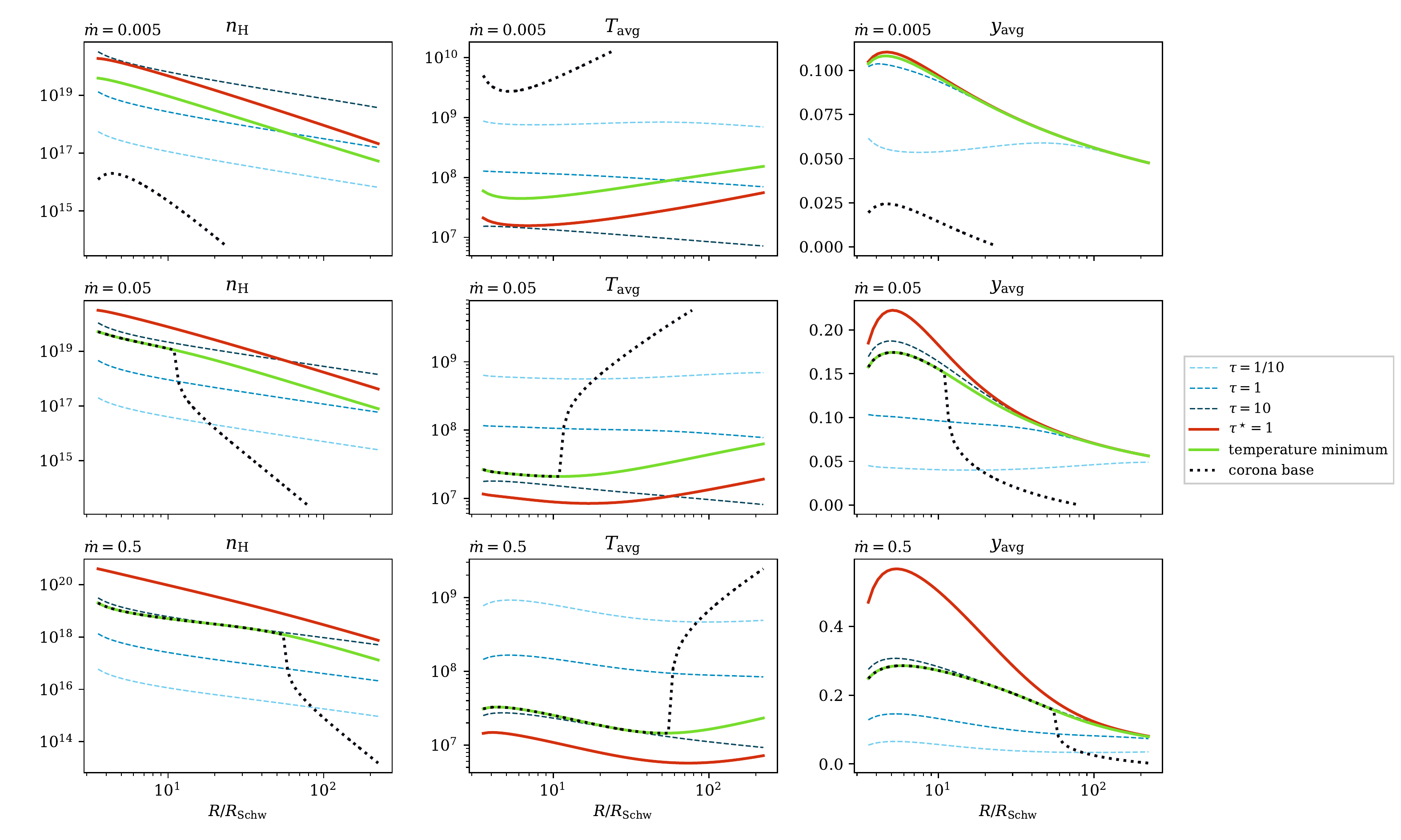}
    \caption{Radial profiles of selected quantities: gas density $\rho$ [cm$^{-3}$], average temperature $T_{\rm avg}$ [K] and average Compton $y_{\rm avg}$ (Eq.~\ref{eq:yavg}), measured at different depths in the disk atmosphere.
   Values at $\tau = 0.1$, $\tau = 1$ and $\tau = 10$ are showed by dashed blue lines, at temperature minimum ($\tau_{\rm cor}$) by a green line, at thermalization depth -- red line and at corona base (determined by temperature minimum or thermal instability in case it is developed) with black dotted line.}
   \label{fig:last}
\end{figure*}

To analyze the global behavior of our model within the parameter space, we below define
several diagnostic quantities as proxies for various properties of the disk/corona system.
We intended to select these quantities so they can be compared to observables.
Three figure maps~\ref{fig:maps-MR}, \ref{fig:maps-AR} and \ref{fig:maps-MA} present these
diagnostics depending on distance from the black hole $R / R_{\rm schw}$, accretion
rate $\dot m$ and $\alpha_{\rm B}$ parameter.

Similarly to RMB15, and opposite to most models, our corona does not have a constant
temperature, but it changes with the optical depth, reaching temperature minimum at
$\tau_{\rm cor}$. From X-ray spectral fitting we measure 
some averaged temperature of the warm medium cooled by Comptonization, therefore 
it will be useful to define the average temperature of the corona as
\begin{equation}\label{eq:tavg}
T_{\rm avg} = \tau^{-1} \int_0^{\tau_{\rm cor}} T d\tau^\prime \ 
\end{equation}
in the aim to compare it with observations. Both quantities $T_{\rm avg}$ and $\tau_{\rm cor}$
 are shown at each panel  a) of three figure maps.  
 Warm, optically thick corona tends to form above the inner strongly magnetized disk of
 high accretion rate.

 Analogously to \cite{1991-HaardtMaraschi}, we denote the fraction of thermal energy released
 in the corona (versus total thermal energy) as
\begin{equation}\label{eq:chipar}
f = \frac{F_{\rm rad}^{\rm cor}}{F_{\rm rad}^{\rm tot}} = 1 - \frac{F_{\rm rad}^{\rm disk}}{F_{\rm rad}^{\rm tot}} \ .
\end{equation}
$f = 0$ corresponds to the passive corona whereas $f = 1$ corresponds to passive disk.   
This parameter is shown in panel d) and it is tightly correlated with magnetic field
gradient $q$ (Eq.~\ref{eq:xcor}), plotted with contours.
Since $f$ is dependent only on magnetic parameters, we suggest that strong magnetic field is
required to obtain high values of $f$ as reported by observations.
Although we never approach $f \approx 1$, thus cannot reproduce the passive disk scenario
discussed by \cite{2013-Petrucci-Mrk509,2017-Petrucci} for the case of AGN, we show that up
to half of total accretion energy can be dissipated in the corona.

Another useful quantity is Compton $y$ parameter, which can also be determined from observations.
To derive the $y$ value for our corona, we use the following definition, which incorporates
an extra term accounting for optically thick medium
\begin{equation}\label{eq:yavg}
y_{\rm avg} = \int_0^{\tau_{\rm es}} \frac{4 k (T - T_{\rm rad})}{m_e c^2} \left( 1 + 2 \tau_{\rm es}^\prime \right) d\tau_{\rm es}^\prime .
\end{equation}
The value of the above parameter is shown in panels e) of Figs.~\ref{fig:maps-MR}, \ref{fig:maps-AR} and \ref{fig:maps-MA} and can range between 0 and 0.5.
This is consistent with values obtained from observations of X-ray binaries: 
$y \approx 0.3$ \citep{2001-Zycki},
$y \approx 0.4$ \citep{2006-Goad}, 
up to $y \sim 1$ \citep{2000-Zhang}.

All above quantities ($\tau_{\rm cor}$, $T_{\rm avg}$,  $f$ and $y_{\rm avg}$) positively
correlate both with magnetic field strength and with released accretion power. 
It means that the most prominent corona would be observed in a case of high accretion rate and
strong field. However, above some threshold of disk magnetization (around $\alpha_{\rm B} = 0.2$
which corresponds to $q \approx 1.3$) the trend reverses, and magnetic flux escapes without
releasing all of its energy into radiation, as it is shown in panel i) of Figs.~\ref{fig:maps-AR}
and \ref{fig:maps-MA}.

In the case of GBHBs, where disk temperatures are generally high, the optical depth of corona 
$\tau_{\rm cor}$ is not usually limited by thermalization layer $\tau^\star = 1$, since the medium
is strongly scattering dominated, as presented in  panel f) of Fig.~\ref{fig:maps-MR}. 
The electron scattering optical depth of that layer is usually ten times larger than
$\tau_{\rm cor}$ for the parameters considered here. Photons are getting thermalized due to high
free-free absorption as in comparison to the total Planck opacity is demonstrated at the same
panel. 

Even if the corona can potentially exist as a solution, thermal instability will occur
in many cases, as discussed in Sec.~\ref{sec:instabil} and \ref{sec:loc}.
Panels g) of Figs.~\ref{fig:maps-MA},n\ref{fig:maps-AR} and \ref{fig:maps-MA}  show the
stability of the worm corona solutions.
The dependence on accretion power is very abrupt and almost switch-like: if the matter gets
too cold or too dense, thus the contribution of bremsstrahlung cooling increases, the thermal
instability occurs, and the corona cannot exist as a stable, uniform layer (regions marked with
magenta. The same effect is obtained when the magnetic field strength increases and the
disk gets colder (compare with Fig.~\ref{fig:multi-A-tau}), but the transition is
less sensitive and more gradual.
The instability appears mostly in cold disk regions, i.e. for low accretion rate and
high magnetic field, and will remove only the optically thick skin when bremsstrahlung
contribution is significant, most likely not affecting the optically thin, hot corona,
which still will be present, even if its optical depth may vary with radius and accretion rate.
This result, particularly the existence of the corona only around the zone of maximum
energy dissipation rate (around $R / R_{\rm schw} = 10$) is consistent with
\cite{1994-KusunoseMineshige}.

Disk thickness $d = H/R$, where $H$ is defined by Eq.~\ref{eq:hdisk} is shown in panels~i)
of each figure maps. It ranges between $10^{-3}$ and $10^{-1}$, which means we still
remain in thin disk regime, consistent with the assumption.
It is important to notice that the disk thickness changes mostly with the accretion rate.
The increase in magnetic field strength does not inflate the disk, (Fig.~\ref{fig:maps-MA}, i)
when the radiative pressure is included, as the fraction of energy dissipated within the
disk decreases (panel d), and radiative pressure supported disk core
(as in classical $\alpha$-disk) is not present, as shown by radiative to gas
pressure ratio $\Xi$, at panel~b). Similarly, total column density $\Sigma$, shown by
contours in panels~c) is decreased, as magnetic fields takes over the role of supporting
the disk structure.


\section{Discussion and Conclusions}
\label{sec:dis}

Disk with optically thick corona is a frequently used satisfactory representation of the 
accretion sources in the soft state, and the present paper is devoted to self-consistent modeling of such a structure, based on first principles.
Using semi-analytical, static model of the magnetically supported disk (MSD) with radiation,
we developed new relaxation numerical code to calculate disk/corona transition. 
The relaxation method designed by us and described in Sec.~\ref{appen1}
is general \citep{Henyey1964} and solves any set of 
linear differential equations with assumed boundary conditions. For the purpose of an accretion
disk with corona we solve the set of five equations, four of them being differential and one algebraic, with four boundary
conditions (Sec.~\ref{eq:boundary}). 

Our code fully solves the vertical structure, taking into account that part of the 
magnetic energy generated in MRI process is converted into radiation and can be dissipated
in the warm, Compton cooled corona. The remaining part of magnetic flux is channeled
into toroidal magnetic field and escapes from the system as introduced by BAR15.
But the original model proposed by BAR15 included no energy balance checking or radiative
transfer, which allowed reducing the model to dimensionless equations. 
In this paper, we combined MSD of those authors with solving 
the radiation transfer  through the gray atmosphere in radiative and hydrostatic 
equilibrium. Additional  mean opacities due to free-free absorption are taken into account.

In case of GBHBs with black hole of stellar mass, we
were able to reproduce both optically thick and thin corona, which is commonly 
detectable in observations of accreting sources. 
Contrary to our previous paper (RMB15), where the corona heating was artificially assumed 
and kept constant with height,
here we obtained self consistent transition through all three layers: 
(i) deep accretion disk up to a few scale heights which is dense and fully thermalized,
(ii) warm corona, where magnetic pressure slows the density decrease and temperature starts 
slowly to increase, (iii) hot Corona where the density is low.
Compton cooling is fully taken into account in all layers, but it strives against
free-free cooling depending on the local physical gas conditions (see panel e) in
Figs.~\ref{fig:multi-M-tau} and ~\ref{fig:multi-A-tau}).

The base of the corona at temperature minimum $\tau_{\rm cor}$ may reach 10 electron scattering
optical depths, while the thermalization zone can be even 60 times $\tau_{\rm es}$
(panels a) and f) in Figs.~\ref{fig:maps-MR}, \ref{fig:maps-AR} and \ref{fig:maps-MA}). 
This fact is in agreement with some observations of warm Compton thick corona by
\cite{2000-Zhang,2001-Zycki}.
The temperature, optical depth and energetic output of the corona are dependent on distance
from the black hole, accretion rate and toroidal magnetic field strength.
High accretion rate and strong toroidal field are favorable conditions
for development of warm $\sim 5$~keV and optically thick $\tau_{\rm cor} \sim 11$ corona,
as seen at the upper left corner of panel a) in Fig.~\ref{fig:maps-MA} . 

Magnetic field strength, $\beta_{0}$, seems to have a leading role in regulating the formation of the corona
since the energy is stored very efficiently in the poloidal magnetic field near the equatorial
plane, and the Poynting flux, $F_{\rm mag}$, quickly increases, reaching its maximum value
at approximately $z / H \sim 2$ for the case presented in Fig.~\ref{fig:cute} panel d).
At this point, the magnetic energy stored in the outflowing field is gradually released by
heating the gas. When going outward an atmosphere, this heating rate overtakes the
magnetic energy gain from the toroidal field production (MRI). 
Asymptotically, the Poynting flux will fully convert to the radiation flux,
but practically this convergence may be very slow in some cases (particularly, for strongly
magnetized disks, where the magnetic pressure declines slowly with height).
Therefore, the magnetic pressure gradient $q$ (Eq.~\ref{eq:qcor}) is an important quantity,
influencing the magnetic energy release and, in consequence, formation of the corona
(see panel b) in Figs.~\ref{fig:multi-M-tau} and \ref{fig:multi-A-tau}).
Stronger toroidal field i.e. lower $\beta_{0}$, implies less steep
gradient and more accretion energy transported from the disk to the atmosphere by Poynting flux.
This fact is nicely demonstrated on at panel b) of Fig.~\ref{fig:maps-MA}, where the
values of $\beta_{0}$ are the smallest, where the warm corona has the largest optical thickness
(panel a) of the same figure). 
On the other hand, the radiative flux $F_{\rm rad}$ will never saturate asymptotically to a
constant value for $q \le 1$ (see panels d) and i) in Fig.~\ref{fig:maps-MA}).

However, optically thick corona may be a subject to local thermal instability caused by
contribution of free-free opacity in the transition layer. Such instabilities were problematic 
in radiative transfer computations of an accretion disk atmospheres of AGN for decades
\citep{krolik1995,2000-Nayakshin,2002-Rozanska-1,2015-Rozanska}. We were expecting that
in case of hotter disks around black holes of stellar mass those instabilities will not
be present, and in addition magnetic heating may remove them. But our results show the opposite,
stronger field expands
corona vertically and at the same time prevents formation of a very hot, radiation
pressure supported disk, which might be sensitive to thermal instabilities. 
When it happens, two outcomes are possible: 
the sharp transition between cold and hot phase will be smoothed out by thermal
conduction and form a relatively plain and regular atmosphere
\citep{1999-Rozanska-Conduction}, or prominence-like clumps of dense matter will form,
and will coexist at the same level with hot medium, similar to lower solar corona.
The clumps could condense or simply emerge from the disk pulled by magnetic field buoyancy
\citep[Fig. 1 in ][]{2014-Jiang}.
Given the violent and turbulent nature of assumed engine powering the disk (MRI),
the clumpy two-phase medium interpretation seems more likely, but also poses a huge
challenge for spectrum modeling.
Variations of this scenario, often considering not only corona but entire disk being clumpy,
has been discussed in the literature \citep{2010-Schnittman-Clumpy,2015-Yang-Clumpy,2016-Wu-Clumpy}.
Regardless of the interpretation, we show that thermal instability occurs in vast amount of
cases, even though GBHB disks are hot and dense and not as prone to it as AGN disks.

Despite the fact that thermal instability affects the very same transition region that is 
responsible for producing optically thick corona,  our model puts constrains on the parameters
for which warm or hot corona can exist. 
For this purpose, we draw the radial extension of the corona base determined by the usual
temperature minimum in case where thermal instability is not present, or by the temperature
of upper stable brunch where thermal instability starts to developed (cross point at upper
pannels of Fig.~\ref{fig:instabil}) by black dotted line in Fig.~\ref{fig:last}.
Gas density, average temperature and Compton parameter radial profiles at selected optical
depths for three accretion rates are shown in Fig.~\ref{fig:last}.

When accretion rate is low, thermal instability arises at each distance from the black hole and 
only hot corona appears, which is four orders of magnitude less dense than the disk atmosphere
(three upper panels of Fig.~\ref{fig:last}). Such corona would be optically thin and the transition 
between disk and corona occurs at about $z/H \sim20$ as demonstrated in  Fig.~\ref{fig:instabil}
upper panels.
 For $\dot m=0.05$, the  warm stable corona can form up to 
 $R \sim 10 \, R_{\rm Schw}$, and such corona is dense and optically thick reaching at the base
 almost $\tau_{\rm es} \sim 10$ (three middle panels of Fig.~\ref{fig:last}).
 Nevertheless, further out from the
 black hole, thermal instability arises and stable corona is again very hot and rare.
 Such hot and rare outer corona dissipates only about one percent of local energy and cannot 
 be associated with usual hard X-ray corona observed in accreting sources. 
Finally, for high accretion rate  $\dot m=0.5$ the radial extent of the warm corona is 
the largest $\sim 60 \, R_{\rm Schw}$ and optical depth $\tau_{\rm es}=10$. Thermal instability and 
hot stable corona appear further away from the black hole. 
 Our results strongly indicate
that the  warm optically thick corona tends to develop in the innermost disk regions,
while outer disk is covered only by rare hot corona. Furthermore, the warm corona
is more radially extended for disks of high accretion rates.
This feature explains why we see soft X-ray excess only in most luminous accreting
sources.

Our model predicts corona formation which is adjusted self consistently by model parameters,
therefore it puts constrains on several quantities which can be directly comparable
with observations. The fist straightforward outcome is the value of temperature
$T_{\rm avg}$ being in the range from 0.1-10 keV.
The second constrained parameter is the density of the warm corona which arises from the dense disk 
atmosphere. The predicted density from our model is of the order of $10^{16-20}$~cm$^{-3}$.  Such value of density is very high, but it agrees with the density of the ionized outflow in accreting black holes 
determined from observations by \cite{2014-Miller}. Our model supports the scenario that eventual disk 
wind can arise from the upper dense layers of magnetically driven accretion disk atmosphere. 

The most important observable is the ratio of thermal energy released in the corona $f$ in comparison 
to total energy
generated in the system. Such ratio is always larger than zero since we compute
Compton cooled atmosphere, which is characterized by typical temperature inversion with
the increasing optical depth  \citep{rozanska2011}. In the frame of our model,  we show that up
to half of total accretion energy can be dissipated in the corona in case of GBHBs
(up to f=0.75 for maximal value of $\alpha_{\rm B}$), therefore we are not able to reproduce 
passive disk scenario discussed by  \cite{2013-Petrucci-Mrk509,2017-Petrucci} for the case of AGN. 
This is natural consequence of our assumed mechanism of heating, which although can transport 
the large portion of energy to the corona, it requires the very effective magnetic coupling with 
gas down to the equatorial plane.  

The value of Compton parameter averaged over optical depth of the warm corona, determined 
from our model is consistent with values determined from X-ray observations  of 
several GBHBs: $y \approx 0.3$ \citep{2001-Zycki},
$y \approx 0.4$ \citep{2006-Goad}, 
up to $y \sim 1$ \citep{2000-Zhang}.

One of the main postulates of BAR15 model is that it can explain the X-ray binary spectral
state transition observed in many sources \citep[e.g.][]{2013-Yamada}. 
by smooth transition from soft to hard state by introducing and intermediate dead zone between
disk and coronal accretion flow.
They utilized the condition for MRI shutoff when gas energy is not enough to sustain the
dynamo \citep{2005-PessahPsaltis}.
In our model, we consider only one zone, where MRI is always operating. We suggest here
that the spectral state transition should occur depending on the strength and
radial extension of the warm, optically thick corona rather than the
full MRI shutoff. Note, that in the frame of our model we do not consider 
the transition from optically thick disk to inner ADAF (advection dominated accretion flow). 

As discussed in BAR15 and SSAB16, one of the key features of MSDs is that they require strong,
global poloidal magnetic field, in order for magnetic dynamo to operate and generate the
toroidal field that actually supports the disk. In case of our typical model $\alpha_{\rm B}=0.1$,
the toroidal magnetic field on equatorial plane and at the base of corona typically
is of the order of $10^8$~G in the inner disk region and of $10^5$~G for radii above
$70\,R_{\rm Schw}$. Those values are smaller than those detected in pulsars, but
are fully in agreement with the prediction of poloidal magnetic filed value required for the jet
launching mechanism. 

Studies suggest that the presence of poloidal magnetic fields is important in jet launching
mechanisms \citep[see also][]{2009-Beckwith-Jets,2012-McKinney-Jets,2018-Liska-Jets}.
\cite{2000-Zhang} noted that for both objects where the optically thick corona was observed,
jets were also present.
Our study shows that the most prominent warm corona will occur in the state when dynamo is
very effective (high $\alpha_{\rm B}$) and magnetic field contribution to disk structure is
large (low $\beta_0$).
On the other hand, no jet was observed in Mrk~509 \citep{2013-Petrucci-Mrk509}.
Nevertheless, the case of AGN will be presented in the forthcoming paper.

\begin{appendix}
\section{Numerical procedure} \label{appen1}

The solution of the set of 5 equations with 4 boundary conditions described in previous 
sections allows determine the density $\rho$, gas temperature $T_{\rm gas}$, radiation
pressure $P_{\rm rad}$, radiative flux $F_{\rm rad}$, and magnetic pressure $P_{\rm mag}$ at
each point of the vertical profile.  The requirement to satisfy 4 boundary conditions,
2 on each side of the computational interval, requires multiple iterations when using
shooting methods, such as Runge-Kutta scheme.
Moreover, the equations are highly nonlinear, and keeping the integration error low requires
high number of grid points (up to $10^5$), even using 4th order Runge-Kutta scheme.
Although computation time using shooting method is still below a few minutes, it becomes a
problem when investigating the global behavior over the parameter space, which requires
computation of thousands of models for different disk and magnetic parameters.
Therefore, to combat this issue we developed a variant of relaxation method, originally described by \cite{Henyey1964} for stellar evolution problems.

The relaxation methods are iterative, when iterations undergoes for the functions defined 
in the whole space of arguments. In the first step,  
some approximate solution for the functions, which is possibly close to the 
exact solution is needed. Then by applying the procedure described below, the correction to entire profile is obtained and added to the original solution.
Since we have many strongly nonlinear terms in our equations, while the method is intrinsically linear, we apply a reduction factor for these corrections in the first couple of iterations to avoid overshooting.
The new solution is then used as an input to the next step and the iteration continues, until the mean relative value of the obtained correction falls below $10^{-7}$, which we consider a convergence criterion.

To apply the relaxation method to our problem, define a grid of $N$ points, in each we have 5 physical quantities. Let us introduce the following notation: the quantity $j = 1 \dots 5$ in point $i = 1 \dots N$ will be denoted $X_i^j$.
We can as well represent it as a vector with 5 components:
\begin{equation}
\vec X_{i} \equiv \left[ \rho_i, T^{\rm gas}_i, T^{\rm rad}_i, F^{\rm rad}_i, P^{\rm mag}_i \right].
\end{equation}
Than we discretize the set of all 5 equations together with boundary conditions. 
For non-differential equations (balance equation and boundary conditions), we evaluate them at respective grid points i.e. $ X^j(z) \rightarrow X^j_{i}$.
For all 4 differential equations we replace quantities with midpoint values and their spatial derivatives with finite differences.
As they are evaluated at midpoints, there are only $N-1$ terms.
\begin{align}
X^j(z) &\rightarrow \frac{X^j_{i} + X^j_{i+1}}{2} \equiv X^j_{i+1/2}, \\
\left. \frac{dX^j}{dz} \right|_{z} &\rightarrow \frac{X^j_{i+1} - X^j_{i}}{z_{i+1} - z_{i}} \equiv Y^j_{i+1/2}.
\end{align}

The core of the problem is to perturb the entire profile to obtain a new one, closer to the solution.  Using the chain rule, we can express the computed profile in terms of perturbations to actual physical quantities:
\begin{equation}\label{eq:chain1}
\delta A^k_i 
= \sum_{j=1}^5 \sum_{l=1}^N \frac{\partial A^k_i}{\partial X^j_l} \delta X^j_l
= - A^k_i,
\end{equation}
where $ i=1 \dots N-1 $ and $ k=1 \dots 5$.
Most of the terms in this sum is zero, as equations evaluated at point $i$ depend only on quantities on points $i$ and $i+1$, thus $l \in \{i,i+1\}$.  
The goal of relaxation method is to transform all equations 
where $A_i^k = 0$, which means that each of the equations $k = 1 \dots 5$ is satisfied in each point $i = 1 \dots N$.
However, having started from some initial guess, this is more likely not true, and the remainder $A_i^k \ne 0$.

Analogically, we do the same for boundary conditions. If we denote the remainder of lower boundary conditions as $L$ and upper boundary conditions as $U$, the expansions in 
terms of perturbations with respect to variables yield
\begin{gather}
\delta L^k 
= \sum_{j=1}^5 \frac{\partial L^k}{\partial X^j_1} \delta X^j_1
= - L^k \quad \text{where} \ k=1,2 ,
\label{eq:chain2}
\\
\delta U^k 
= \sum_{j=1}^5 \frac{\partial U^k}{\partial X^j_N} \delta X^j_N
= - U^k \quad \text{where} \ k=1,2 .
\label{eq:chain3}
\end{gather}

At this point we have mathematically well defined problem: we must solve the equations \eqref{eq:chain1}, \eqref{eq:chain2} and \eqref{eq:chain3} to obtain the profile corrections $\delta X^{1 \dots 5}_{1 \dots N}$.
We can express it as perturbation vector ${\bf \delta X}$, remainder vector ${\bf R}$ and Jacobian matrix ${\bf M}$:
\begin{equation}\label{eq:linearsystem}
{\bf M} \cdot {\bf \delta X} = - {\bf R} 
\end{equation}

We arrange our 5 variables and 5 equations into a vector using following layout:
\begin{align}
X_{5i - 5 + j} &= X_i^j \quad \text{where} \  j = 1 \dots 5  \  \text{and}  \  i = 1 \dots N  ,
\\
R_{k} &= L^k \quad \text{where} \  k = 1,2  ,
\\
R_{5i - 3  + k} &= A^k_i \quad \text{where} \  k = 1 \dots 5  \  \text{and}  \  i = 1 \dots N  ,
\\
R_{5N - 2  + k} &= U^k  \quad \text{where} \  k = 1,2  .
\end{align}

Now the remaining issue is to fill the matrix accordingly.
For linear equations,  to  evaluate all nonzero matrix terms for
$i = 1 \dots N$, $j = 1 \dots 5$, $k = 5$ and $l = i$, we use directly the formula
\begin{equation}\label{eq:matrixfill}
M_{pq}  = \frac{\partial A^k_i}{\partial X^j_l},
\quad \text{where} \  
\begin{array}{l} 
p = 5i - 3 + k \\ 
q = 5l - 5 + j .
\end{array}
\end{equation}
For differential equations, each point has dependence on two consecutive points, and we evaluate these terms for $i = 1 \dots N-1$, $j = 1 \dots 5$, $k = 1 \dots 4$ and $l \in \{i,i+1\}$ by using the chain rule:
\begin{align}
\frac{\partial A^k_i}{\partial X^j_i} 
&= \frac{\partial X^j_{i+1/2}}{\partial X^j_i} \frac{\partial A^k_i}{\partial X^j_{i+1/2}} 
+ \frac{\partial Y^j_{i+1/2}}{\partial X^j_i} \frac{\partial A^k_i}{\partial Y^j_{i+1/2}} \\
&= \frac{1}{2} \frac{\partial A^k_i}{\partial X^j_{i+1/2}} 
- \frac{1}{z_{i+1}-z_i} \frac{\partial A^k_i}{\partial Y^j_{i+1/2}},
\\
\frac{\partial A^k_i}{\partial X^j_{i+1}} &= \frac{\partial X^j_{i+1/2}}{\partial X^j_{i+1}} \frac{\partial A^k_i}{\partial X^j_{i+1/2}} 
+ \frac{\partial Y^j_{i+1/2}}{\partial X^j_{i+1}} \frac{\partial A^k_i}{\partial Y^j_{i+1/2}} \\
&= \frac{1}{2} \frac{\partial A^k_i}{\partial X^j_{i+1/2}} 
+ \frac{1}{z_{i+1}-z_i} \frac{\partial A^k_i}{\partial Y^j_{i+1/2}}.
\end{align}
Finally, for boundary conditions, we evaluate the matrix coefficients at lower ($L$) and upper ($U$) boundary as
\begin{gather}
M_{kj} = \frac{\partial L^k}{\partial X^j_1}
\quad \text{where} \quad k = 1,2,  \label{eq:matrixfill2} 
\\
M_{pq} = \frac{\partial U^k}{\partial X^j_N}
\quad \text{where} \quad 
\begin{array}{l} 
k = 1,2		   \\
p = 5N - 2 + k  \\ 
q = 5N - 5 + j .
\end{array} \label{eq:matrixfill3}
\end{gather}



The relaxation method provides excellent precision, even for low number of computational points. Because the method is adjusting the entire profile at once, allowing us to investigate even marginally stable cases. The method is also clearly expandable for higher order derivatives.

Its major disadvantage is a requirement to evaluate the partial derivatives of left-hand sides of all equations (and boundary conditions) with respect to all variables and their spatial derivatives.
This makes the code very difficult to modify and error prone.
We handle this issue by utilizing symbolic computation techniques with {\it sympy} framework \citep{2017-Meurer-sympy}, which also allows to generate ready-to-compile procedures to evaluate the matrix coefficients that we can include in our code written in Fortran 2008.
We then solve the system of linear equations using {\tt DGBSV} procedure from {\it LAPACK} \citep{1990-Anderson-LAPACK} and obtain corrections to all 5 variables at each point.
This method is very fast and allows us to obtain a precise solution in less than half of second on typical desktop computer. \nocite{2011-Tange-GNUparallel}
 \end{appendix}

\section*{Acknowledgments}
We thank Bo{\.z}ena Czerny for all comments and helpful discussion.  
This research was supported by  Polish National Science Center grants No.
 2015/17/B/ST9/03422 and 2015/18/M/ST9/00541.


\bibliographystyle{aa}
\bibliography{refs,general,disks,tools}

\end{document}